\documentclass[epj,nopacs]{svjour}

\hugehead

\DeclareMathAlphabet{\mathpzc}{OT1}{pzc}{m}{it}
\usepackage{lineno}

\usepackage{graphicx}
\usepackage{verbatim}
\usepackage[percent]{overpic}
\usepackage{float}
\usepackage{array}
\usepackage{multirow}
\usepackage{epsfig}
\usepackage{color}

\newcommand{\minitab}[2][1]{\begin{tabular}{#1}#2\end{tabular}}

\newcommand{\aaa}{1}
\newcommand{\bbb}{2}
\newcommand{\ccc}{3}
\newcommand{\ddd}{4}
\newcommand{\eee}{5}
\newcommand{\fff}{6}
\newcommand{\ggg}{7}
\newcommand{\hhh}{8}
\newcommand{\iii}{9}
\newcommand{\jjj}{10}
\newcommand{\kkk}{11}
\newcommand{\lll}{12}
\newcommand{\mmm}{13}
\newcommand{\nnn}{14}
\newcommand{\ooo}{15}
\newcommand{\ppp}{16}
\newcommand{\qqq}{17}
\newcommand{\rrr}{18}
\newcommand{\sss}{19}
\newcommand{\ttt}{20}
\newcommand{\uuu}{21}
\newcommand{\vvv}{22}
\newcommand{\www}{23}
\newcommand{\xxx}{24}
\newcommand{\yyy}{25}
\newcommand{\zzz}{26}
\newcommand{\AAA}{27}
\newcommand{\BBB}{28}
\newcommand{\CCC}{29}
\newcommand{\DDD}{30}
\newcommand{\EEE}{31}
\newcommand{\FFF}{32}
\newcommand{\GGG}{33}
\newcommand{\HHH}{34}
\newcommand{\III}{35}

\begin{document}

\title{Integral measurement of the $^{12}$C(n,p)$^{12}$B reaction up to 10~GeV}

\author{
P.~\v{Z}ugec\inst{\aaa,}\thanks{\email{pzugec@phy.hr}}
\and
N.~Colonna\inst{\bbb}
\and
D.~Bosnar\inst{\aaa}
\and
A.~Ventura\inst{\ccc}
\and
A.~Mengoni\inst{\ddd}
\and
S.~Altstadt\inst{\eee}
\and
J.~Andrzejewski\inst{\fff}
\and
L.~Audouin\inst{\ggg}
\and
M.~Barbagallo\inst{\bbb}
\and
V.~B\'{e}cares\inst{\hhh}
% ...
}


\institute{
Department of Physics, Faculty of Science, University of Zagreb, Croatia
\and
Istituto Nazionale di Fisica Nucleare, Sezione di Bari, Italy
\and
Istituto Nazionale di Fisica Nucleare, Sezione di Bologna, Italy
\and
ENEA, Bologna, Italy
\and
Johann-Wolfgang-Goethe Universit\"{a}t, Frankfurt, Germany
\and
Uniwersytet \L\'{o}dzki, Lodz, Poland
\and
Centre National de la Recherche Scientifique/IN2P3 - IPN, Orsay, France
\and
Centro de Investigaciones Energeticas Medioambientales y Tecnol\'{o}gicas (CIEMAT), Madrid, Spain
\and
Charles University, Prague, Czech Republic
\and
European Commission JRC, Institute for Reference Materials and Measurements, Retieseweg 111, B-2440 Geel, Belgium
\and
CEA/Saclay - IRFU, Gif-sur-Yvette, France
\and
University of Manchester, Oxford Road, Manchester, UK
\and
CERN, Geneva, Switzerland
\and
Universitat Politecnica de Catalunya, Barcelona, Spain
\and
C2TN-Instituto Superior Tecn\'{i}co, Universidade de Lisboa, Portugal
\and
Universidad de Sevilla, Spain
\and
INFN - Laboratori Nazionali del Sud, Catania, Italy
\and
National Technical University of Athens (NTUA), Greece
\and
Instituto de F{\'{\i}}sica Corpuscular, CSIC-Universidad de Valencia, Spain
\and
Paul Scherrer Institut, 5232 Villigen PSI, Switzerland
\and
Universidade de Santiago de Compostela, Spain
\and
Aristotle University of Thessaloniki, Thessaloniki, Greece
\and
Bhabha Atomic Research Centre (BARC), Mumbai, India
\and
Atominstitut der \"{O}sterreichischen Universit\"{a}ten, Technische Universit\"{a}t Wien, Austria
\and
University of York, Heslington, York, UK
\and
Karlsruhe Institute of Technology (KIT), Institut f\"{u}r Kernphysik, Karlsruhe, Germany
\and
Dipartimento di Fisica, Universit\`a di Bologna, and Sezione INFN di Bologna, Italy
\and
Istituto Nazionale di Fisica Nucleare, Laboratori Nazionali di Legnaro, Italy
\and
Istituto Nazionale di Fisica Nucleare, Sezione di Trieste, Italy
\and
Horia Hulubei National Institute of Physics and Nuclear Engineering - IFIN HH, Bucharest - Magurele, Romania
\and
Dipartimento di Fisica e Astronomia DFA, Universit\`a di Catania and INFN-Laboratori Nazionali del Sud, Catania, Italy
\and
University of Vienna, Faculty of Physics, Austria
\and
Centre for Astrophysics Research, School of Physics, Astronomy and Mathematics, University of Hertfordshire, Hatfield, United Kingdom
\and
Department of Physics, University of Basel, Basel, Switzerland
\and
Research School of Physics and Engineering, Australian National University, ACT 0200, Australia
}

\date{Received: date / Revised version: date}

\abstract{
The integral measurement of the $^{12}$C(n,p)$^{12}$B reaction was performed at the neutron time of flight facility n\_TOF at CERN. The total number of $^{12}$B nuclei produced per neutron pulse of the n\_TOF beam was determined using the activation technique in combination with a time of flight technique. The cross section is integrated over the n\_TOF neutron energy spectrum from reaction threshold at 13.6~MeV to 10~GeV. Having been measured up to 1~GeV on basis of the $^{235}$U(n,f) reaction, the neutron energy spectrum above 200~MeV has been reevaluated due to the recent extension of the cross section reference for this particular reaction, which is otherwise considered a standard up to 200~MeV. The results from the dedicated GEANT4 simulations have been used to evaluate the neutron flux from 1~GeV up to 10~GeV. The experimental results related to the  $^{12}$C(n,p)$^{12}$B reaction are compared with  the evaluated cross sections from major libraries and with the predictions of different GEANT4 models, which mostly underestimate the $^{12}$B production. On the contrary, a good reproduction of the integral cross section derived from measurements is obtained with TALYS-1.6 calculations, with optimized parameters.
\PACS{
{23.40.-s}{beta decay}
\and
{24.10.Lx}{nuclear-reaction models}
\and
{25.40.Kv}{(n,p) reactions}
} 
} 

\maketitle

\section{Introduction}

Neutron induced reactions are important for a variety of research fields, from fundamental Nuclear Physics and Nuclear Astrophysics to applications of nuclear technologies to energy production, nuclear medicine, material characterization, cultural heritage, etc. Neutron induced reactions on light nuclei -- such as Carbon, Oxygen and Nitrogen -- are of particular interest in nuclear medicine due to their high abundance in the human body. Particularly significant in this respect are the reactions leading to the emission of charged particles. Among them, the $^{12}$C(n,p)$^{12}$B reaction -- occurring at neutron energies above the threshold of 13.6~MeV -- may affect the dose distribution in hadrontherapy or conventional radiotherapy if high energy secondary neutrons are produced during the treatment delivery, since the $^{12}$C(n,p)$^{12}$B reaction leads both to the emission of protons and energetic electrons (6.35~MeV average energy) from the decay of $^{12}$B. This reaction is also important for calculations in radiological protection, as well as for the design of shields and collimators at accelerator facilities, spallation neutron sources and fusion material irradiation facilities -- such as MTS and IFMIF -- whose neutron spectrum shows an important tail extending above the threshold of this reaction \cite{fusion}. Finally, the cross section of the $^{12}$C(n,p)$^{12}$B reaction is important for accurately simulating the response of diamond detector to fast neutrons \cite{rebai,pillon}.

\begin{figure}[b!]
\includegraphics[width=1.0\linewidth,keepaspectratio]{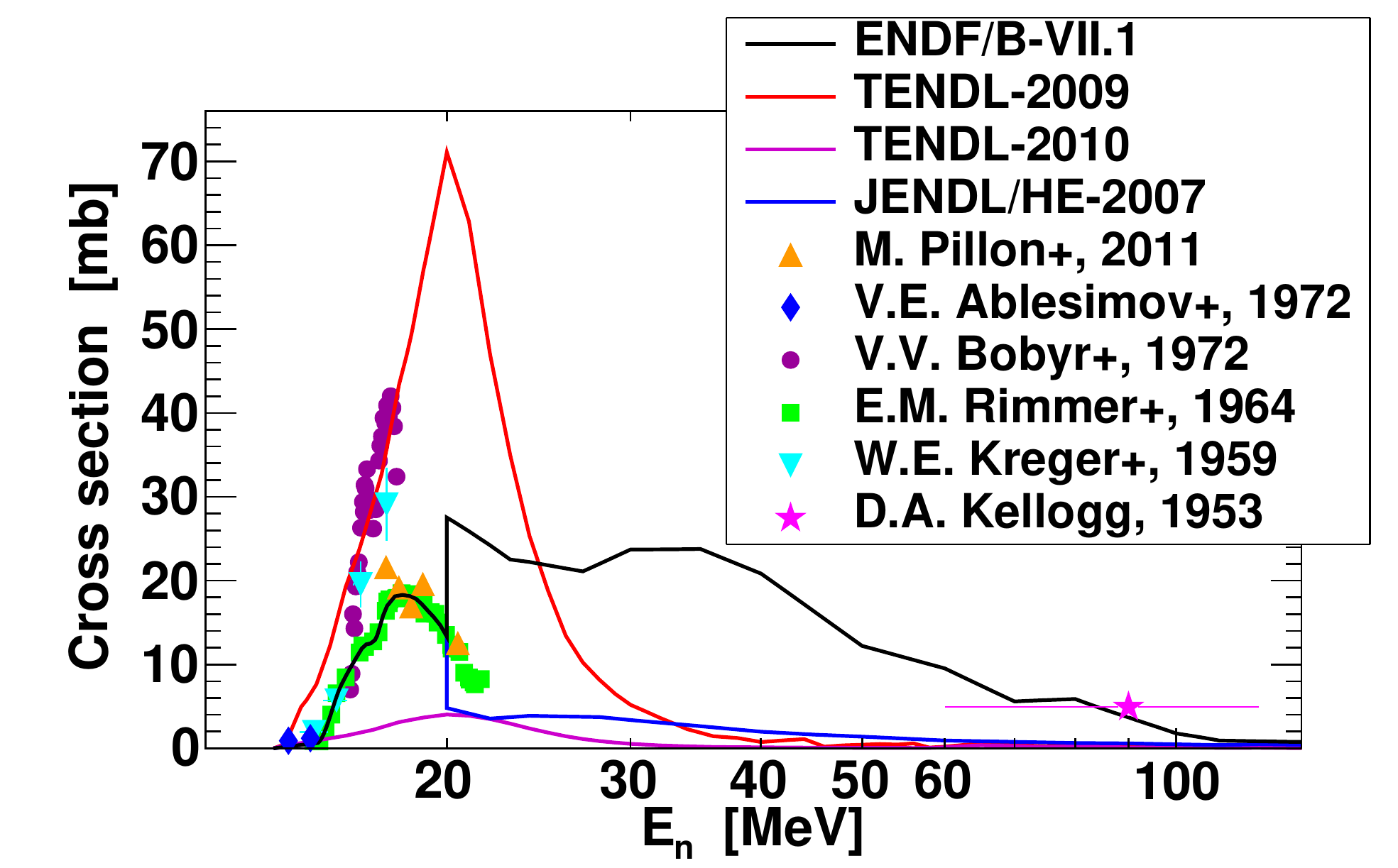}
\caption{Compilation of the available experimental and evaluated data on the $^{12}$C(n,p)$^{12}$B reaction.}
\label{fig0}
\end{figure}

Despite their importance, cross section data for this reaction are scarce and largely discrepant. Figure~\ref{fig0} shows the current status of experimental results. Below 20 MeV four old measurements are present in literature \cite{ablesimov,kreger,rimmer,bobyr}, all obtained with the activation technique, with short pulses of monoenergetic neutrons inducing the reaction, followed by long beam-off intervals for counting the $^{12}$B $\beta$-decay. Recently, Pillon \emph{et al.} measured a series of neutron-induced reactions on carbon for neutron energies from 5 MeV to 20.5 MeV, by using a single crystal diamond detector \cite{pillon}. The measurements were performed at the Van de Graaff neutron generator of the EC-JRC-IRMM, Geel, Belgium. Thanks to the high energy resolution of the device, structures related to various neutron-induced reactions on carbon could be distinguished in the detected energy spectrum. The cross section was extracted from the structures identified as due to the (n,p) reaction. The results, shown in fig.~\ref{fig0}, are consistent with those of ref.~\cite{rimmer}. However, according to the authors, not all possible excited states of the residual nucleus could be identified, thus resulting in an  underestimate of the cross section. Above 20 MeV only one data point from Kellogg \cite{kellog} is present in EXFOR, centered at 90 MeV, but with a poor resolution in energy and a large uncertainty in cross section. 

The discrepancy between existing data below 20 MeV, combined with the scarcity of reliable data above 20 MeV, is reflected in the differences between evaluated cross sections in various libraries, as well as in highly uncertain theoretical predictions at all neutron energies. The cross sections reported in the major libraries are shown in fig.~\ref{fig0}. Up to 20~MeV the latest versions of all libraries contain the same cross section, based purely on the dataset from Rimmer and Fisher \cite{rimmer}. Among older versions, TENDL-2009 predicts a cross section a factor of three higher than all other evaluations. For the reaction considered herein, this library was based on TALYS calculations \cite{talys}. On the other hand, a drastically lower evaluation is adopted by TENDL-2010 and subsequent versions. Above 20~MeV evaluations are based on theoretical estimates. In particular, ENDF/B-VII.1 has adopted calculations performed with the Feshbach-Kerman-Koonin (FKK)-GNASH code described in ref.~\cite{chadwick}, up to 150~MeV \cite{endf}. A completely different cross section, based on calculations by Watanabe \emph{et al.}, \cite{watanabe} is contained in the special high energy file of the Japanese evaluated nuclear data library, JENDL/HE-2007 \cite{jendl,jendl_he}. 

Models can be used in the main Monte Carlo codes for neutron transport to estimate the cross section of this reaction. In particular, the GEANT4 package \cite{geant4} offers various options for calculating the cross section of this reaction, from threshold to several GeV. However, the trends and values of the cross section largely vary, depending on the selected model (as shown in later fig.~\ref{fig2}). While, in principle, these predictions can be checked against experimental data below 20~MeV, with the current status of data discrepancy it is practically impossible to decide which of the various options is more reliable in simulating this particular reaction in GEANT4 or in any other Monte Carlo code for neutron transport. Furthermore, due to the lack of experimental data, nothing can be said about the validity of the calculations at higher neutron energies.

Considering the present status of experimental data, evaluated libraries and model predictions, new accurate data are highly desirable, covering a wide energy range extending from the reaction threshold to the GeV region. Time-of-flight facilities based on the spallation neutron sources could, in principle, be used to this purpose, with an experimental setup able to detect the emitted proton. In practice, however, such a measurement is complicated by the presence of other competing reaction channels, in particular elastic and inelastic scattering, (n,d) and (n,$\alpha$) reactions \cite{rebai}. A somewhat simpler, yet useful approach, would be to perform an integral measurement of the cross section by means of the activation technique with a pulsed neutron beam of a low repetition rate and with an energy spectrum extending much above the reaction threshold. Both requirements are met by the n\_TOF facility at CERN \cite{carlos}, characterized by a white neutron spectrum extending up to $\sim$10~GeV, and a low repetition rate ($\leq$0.8~Hz). A new technique was applied at n\_TOF to extract an energy-integrated cross section. The measurement relied on the detection of the $^{12}$B $\beta$-decay within the same neutron bunch in which the reaction takes place. The main features of the measurement and its results have been reported in ref.~\cite{nprc}. The main aim of the measurement was to provide an information that could be used as a benchmark for validating the evaluated cross sections and the predictions of the model calculations.

In this paper the technique and the analysis procedure are described in greater detail, with the result and its physical implications being discussed at length. The experimental result is compared with various models available in GEANT4, from which the information on the reliability of these models is obtained. Furthermore, it will be shown that the measured observable can be closely reproduced by calculations based on the TALYS code, with optimized parameters.
 
The paper is organized as follows: the experimental setup is presented in Section~\ref{experiment}, while the on-line activation technique and data analysis are discussed in Section~\ref{analysis}. Section~\ref{geant} describes the method used to estimate this cross section from GEANT4 simulations. Section~\ref{integral} is dedicated to the comparison between the experimental result and the predictions of various models and evaluations. A comparison between the various experimental results and the optimized TALYS-1.6 calculations is shown in Section~\ref{theory}. Finally, Section~\ref{conclusion} summarizes the main conclusions of this work.

\section{Experimental setup}
\label{experiment}

The neutron beam at n\_TOF is produced by the proton-induced spallation in the massive Pb target. The pulsed beam of 20 GeV protons is provided by the CERN Proton Synchrotron, delivering an average of $7\times10^{12}$ protons per pulse, with 7 ns spread, a repetition rate in multiples of 1.2~s and a typical average frequency of 0.4~Hz. On average, 300 neutrons are produced per incident proton. Spallation neutrons are moderated passing through the Pb block itself, through 1 cm of demineralized water and 4 cm of borated water surrounding the block. The borated water significantly suppresses the production of 2.2 MeV $\gamma$-rays from neutron capture on hydrogen, thanks to the $^{10}$B(n,$\alpha$)$^{7}$Li reaction. The outgoing neutron flux spans 12 orders of magnitude in energy -- from thermal ($\sim$10~meV) up to $\sim$10 GeV.

An evacuated beam line leads to the Experimental Area 1, at a distance of approximately 185~m from the spallation target. The charged particles are removed from the beam by a 1.5 T sweeping magnet at 145 m from a spallation target, while the neutron beam itself is shaped by a set of two collimators at 137~m and 178~m. Outside the beam line, the flux of particles and radiation from the spallation target is attenuated by massive concrete walls, together with a 3.5~m thick iron shielding. A description of the general features of the n\_TOF facility may be found in ref.~\cite{carlos}.

The neutron flux at n\_TOF is measured by multiple detector systems, in order to reliably cover the whole energy range from thermal up to 1 GeV. In particular, the flux between 10 MeV and 1 GeV, which is the energy range of interest to this work, has been measured by the Parallel Plate Avalanche Counters (PPAC \cite{ppac1,ppac2}) relying on the $^{235}$U(n,f) reaction. The cross section of this reaction -- as the reference for the absolute normalization of the data -- is not considered a standard above 200~MeV. However, the cross section reference for this reaction has recently been extended up to 1~GeV \cite{standard}, and the neutron flux at these energies has been reevaluated accordingly (Section~\ref{flux}). Dedicated GEANT4 simulation of a spallation process were used to extend the flux evaluation up to 10 GeV, with the simulation results normalized to the experimental data around the transition point at 1~GeV. A detailed description of the neutron flux measurements at n\_TOF may be found in ref.~\cite{massimo}.

The natural carbon sample used for the measurement was 7.13~g in mass, 2~cm in diameter and 1~cm in thickness. A high chemical purity of 99.95\% was confirmed by the chemical analysis performed at Paul Scherrer Institute, excluding the possibility of contamination by neutron poisons.

For the detection of $\beta$-rays from a decay of $^{12}$B produced by the $^{12}$C(n,p)$^{12}$B reaction, two C$_6$D$_6$ (deuterated benzene) liquid scintillation detectors were employed, which are commonly used at n\_TOF for neutron capture measurements. The two detectors are referred to as Bicron and FZK. The former is a modified version of the commercially available Bicron detector, while the latter was custom built at Forschungszentrum Karls-ruhe, Germany. These detectors have been specifically optimized so as to exhibit a very low neutron sensitivity \cite{plag}. They were mounted 8.2~cm upstream of the sample and 6.8~cm from the beam line axis. The backward position significantly reduces the background of neutrons and in-beam $\gamma$-rays scattered off the sample. The scintillation liquid volumes amount to 618~ml and 1027~ml for Bicron and FZK, respectively. Their energy calibration was performed using standard $^{137}$Cs, $^{88}$Y and Am/Be $\gamma$-ray sources.

A high-performance digital data acquisition system -- based on 8-bit flash analog-to-digital converter units (FADC) -- was used for recording the electronic signals from the measurements. The sampling rate of 500~MHz in combination with a 48~MB memory buffer allows uninterrupted waveforms of 96~ms duration. Further details on the digital data acquisition used at n\_TOF may be found in ref.~\cite{daq}.

\section{Experimental data analysis}
\label{analysis}

The integral measurement of the cross section of $^{12}$C(n,p)$^{12}$B reaction has been performed using the activation technique in combination with the time of flight technique \cite{nprc}. During the neutron irradiation of the $^\mathrm{nat}$C sample, the high energy section of the neutron beam (above the reaction threshold of 13.6~MeV) causes the production of $^{12}$B nuclei in the sample, which undergo a $\beta^-$-decay with a half-life of 20.2~ms and a $Q$-value of 13.37~MeV. The highly energetic $\beta$-rays -- with a mean kinetic energy of 6.35~MeV -- are then detected by two C$_6$D$_6$ detectors. Though the data have been measured within the time window of 96~ms, due to the reduced statistics after the background subtraction at higher decay times, they have been analyzed only up to 80 ms, which corresponds to four half-lives of $^{12}$B. The intense $\gamma$-flash caused by the proton beam hitting the spallation target was used as a reference point for the time calibration. Due to the low repetition rate of the n\_TOF beam, the decay of $^{12}$B is detected within the same neutron bunch in which it is produced, without any possibility of wrap-around background.

The measurements are affected by several sources of background. One component is caused by the scattering of in-beam $\gamma$-rays off the sample itself. This component was measured with a Pb sample and was found to be negligible. The second component is caused by the neutron beam crossing the experimental area. It was measured by recording data with the beam in the experimental area, but without any sample in place. The third component is the ambient background, caused by the natural radioactivity and the neutron activation. It was measured by turning off the neutron beam. All these components were properly normalized and subtracted from the measured counts with the $^\mathrm{nat}$C sample. These background components have already been discussed in ref.~\cite{58ni}.

The fourth and final component is referred to as the neutron background, caused by the neutrons scattering off the sample itself. To precisely identify this component, one must rely on dedicated simulations, taking into account the detailed geometric description of the detectors and experimental surroundings, the full framework of the neutron induced reactions and their complete temporal evolution. These simulations have been developed in GEANT4 \cite{geant4} and described in detail in ref.~\cite{background}. A major portion of the neutron background is caused by the detection of $\gamma$-rays -- mainly from the neutron captures inside the experimental area -- with only a minor contribution from $\beta$-rays coming from activation of the experimental setup.

A potential problem in the simulations of this background component is the lack of correlations between the capture $\gamma$-rays from simulated $\gamma$-ray cascades, which alters their energy distribution (relative to the real one), thus affecting an average $\gamma$-ray detection efficiency. However, the use of the Pulse Height Weighting Technique (PHWT) \cite{wf} compensates for the lack of correlations, making the detection efficiency independent of the actual cascade paths, as long as the energy is conserved in simulating the capture $\gamma$-rays. In this case simulations become highly reliable, as demonstrated in ref.~\cite{background} by the very good agreement between the simulated and measured yield of $^\mathrm{nat}$C in an energy region where the neutron background dominates. The details on the PWHT applied at n\_TOF may be found in ref.~\cite{wf_ntof}.

After subtracting the experimentally determined background components, the weighted spectrum $C_W(t)$ of counts  per neutron bunch -- as a function of decay time $t$ -- may be expressed as:
\begin{linenomath}\begin{equation}
\label{eq13}
C_W(t)=W(E)\otimes C_\gamma(t)+\langle W\rangle \times C_\beta(t)
\end{equation}\end{linenomath}
Here $W(E)$ is the weighting function from PHWT, dependent on the energy $E$ deposited in the detectors and determined for each detector separately. $C_\gamma(t)$ is the neutron background -- mostly composed of capture $\gamma$-rays -- while $\otimes$ symbolically denotes the operation of applying the weighting function to the neutron background counts. $C_\beta(t)$ are the $\beta$-ray counts, pertaining only to the $\beta$-rays from the decay of $^{12}$B produced in $^\mathrm{nat}$C sample. The $\beta$-rays coming from other sources and reactions -- including the $^{12}$C(n,p)$^{12}$B reaction outside the sample (\emph{e.g.} from the detector housing made of carbon fiber) -- are all part of the neutron background $C_\gamma(t)$. In the case of $\beta$-rays -- always emitted with unit multiplicity -- the application of the PWHT is equivalent to multiplying the overall spectrum by the average weighting factor $\langle W\rangle$, which may be expressed as:
\begin{linenomath}\begin{equation}
\label{eq12}
\langle W\rangle=\frac{\int_{E_\mathrm{min}}^{E_\mathrm{max}}S_\beta(E)W(E)\mathrm{d}E}{\int_{E_\mathrm{min}}^{E_\mathrm{max}}S_\beta(E)\mathrm{d}E}
\end{equation}\end{linenomath}
where $S_\beta(E)$ is the deposited energy spectrum of $\beta$-rays from the $^{12}$B decay. In order to determine the average weighting factor $\langle W\rangle$ from eq.~(\ref{eq12}), the spectrum $S_\beta(E)$ extracted from simulations of the $^{12}$B-decay in the $^\mathrm{nat}$C sample was used (contrary to $\gamma$-ray cascades, there is no difference between the simulated and measured spectrum of the energy deposited by $\beta$-rays in the C$_6$D$_6$). The average value $\langle W\rangle$ depends on the lower and upper thresholds $E_\mathrm{min}$ and $E_\mathrm{max}$ set during the data analysis. In particular, the lower threshold needs to be the same as the threshold used for calculating the weighting functions from PHWT, \emph{i.e.} $E_\mathrm{min}$~=~200~keV. The upper threshold is set to $E_\mathrm{max}$~=~13.37~MeV, which is the \mbox{$Q$-value} of the $^{12}$B decay. The weighted simulated neutron background $C_\gamma(t)$ and the average weighting factor $\langle W\rangle$ -- calculated from the simulated spectrum $S_\beta(E)$ -- are then used to invert eq.~(\ref{eq13}) and determine the background-subtracted, unweighted $\beta$-rays spectrum:
\begin{linenomath}\begin{equation}
C_\beta(t)=\frac{C_W(t)-W(E)\otimes C_\gamma(t)}{\langle W\rangle}
\end{equation}\end{linenomath}
The remaining spectrum corresponds to the time distribution of $^{12}$B decays:
\begin{linenomath}\begin{equation}
\label{eq1}
C_\beta(t)=\frac{\varepsilon_\beta N_{^{12}\mathrm{B}}}{\tau}e^{-t/\tau}
\end{equation}\end{linenomath}
with $\tau=29.14$~ms as the lifetime of $^{12}$B and $ N_{^{12}\mathrm{B}}$ as the total number of $^{12}$B nuclei produced per single neutron bunch. A final factor to be considered is the total $\beta$-ray detection efficiency $\varepsilon_\beta$. It was extracted from simulations, as the ratio between the number of detected $\beta$-rays within the energy thresholds used in the analysis ($E_\mathrm{min}$~=~200~keV and $E_\mathrm{max}$~=~13.37~MeV) and the total number of $^{12}$B nuclei generated in the $^\mathrm{nat}$C sample. For the Bicron detector it was determined as $\varepsilon_{\beta;\mathrm{Bicron}}=4.3\%$, while for the FZK as $\varepsilon_{\beta;\mathrm{FZK}}=6.8\%$.

A fit to the exponential form of eq.~(\ref{eq1}) -- with $N_{^{12}\mathrm{B}}$ as the only free parameter -- yields the total number of $^{12}$B nuclei produced per neutron bunch. For the Bicron detector the fit yields $N^{(\mathrm{Bicron})}_{^{12}\mathrm{B}}=68.03\pm0.66$, while the data from FZK detector yield $N^{(\mathrm{FZK})}_{^{12}\mathrm{B}}=68.74\pm0.44$. 

\begin{figure}[t!]
\includegraphics[angle=90,width=1.0\linewidth,keepaspectratio]{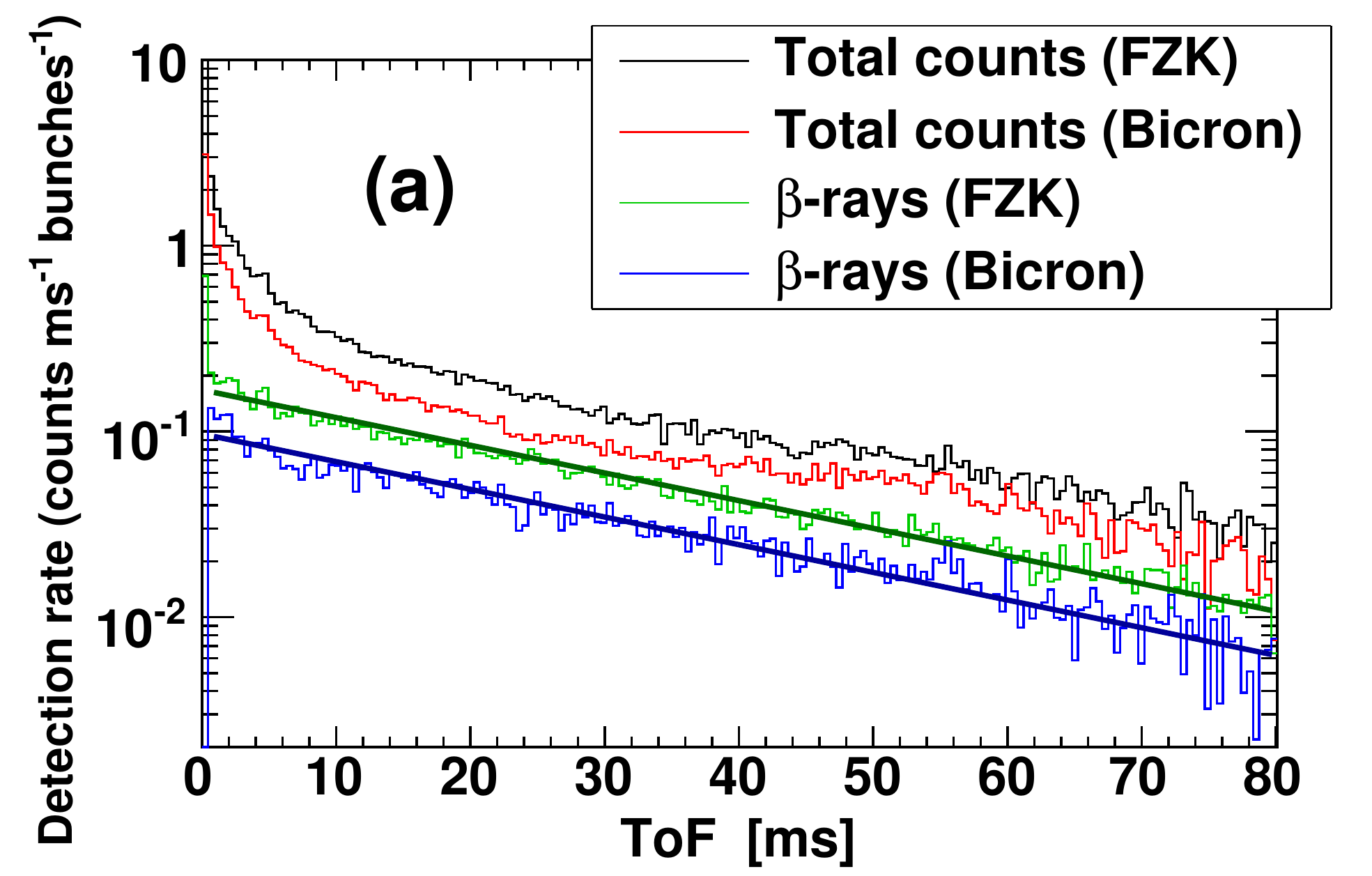}
\includegraphics[angle=90,width=1.0\linewidth,keepaspectratio]{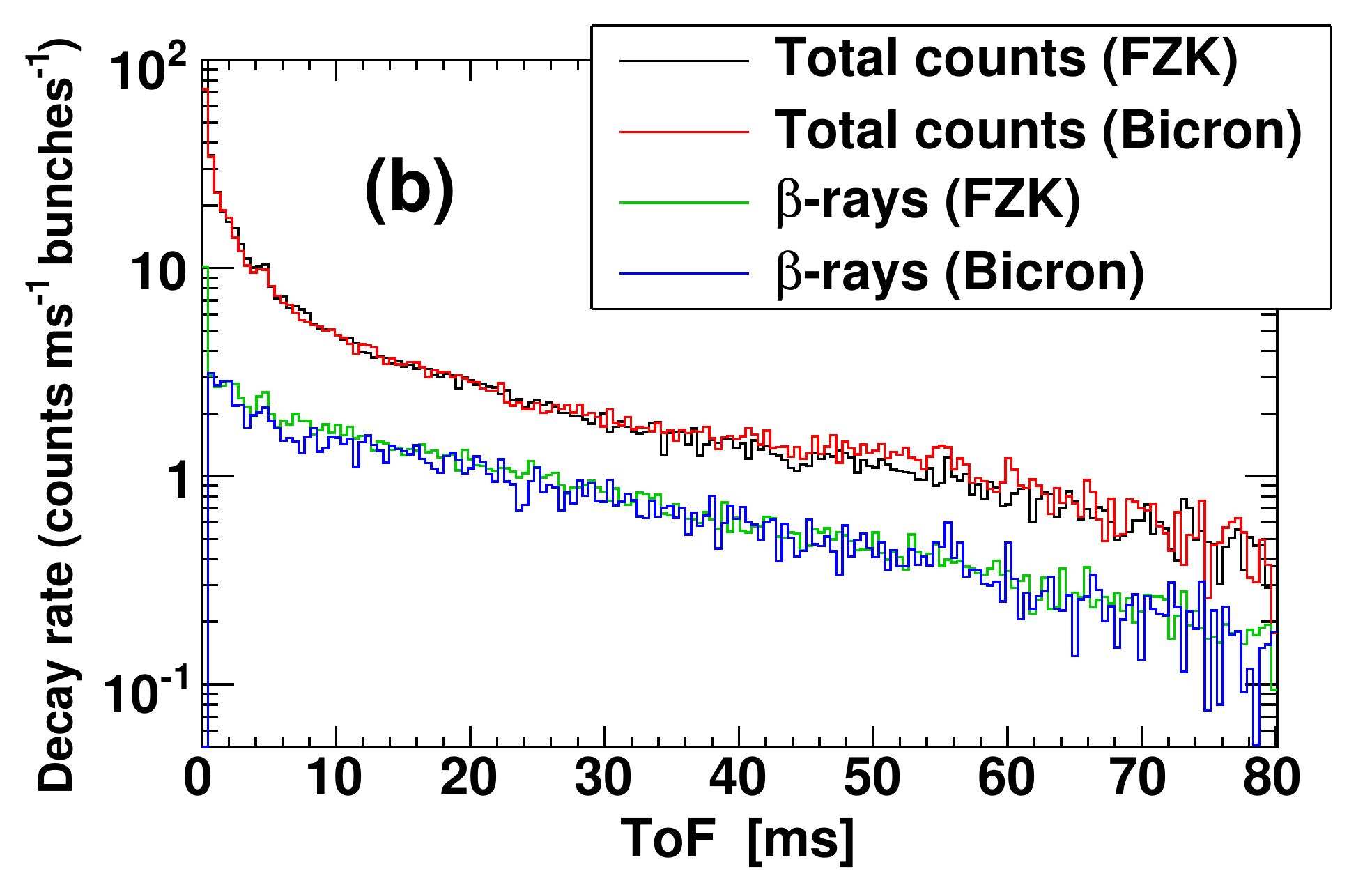}
\caption{Time distribution of the $\beta$-decay of $^{12}$B nuclei produced by the $^{12}$C(n,p)$^{12}$B reaction, measured by two C$_6$D$_6$ detectors (Bicron and FZK). Total counts show the data after subtracting all experimentally identified background components, with the neutron background still remaining. After subtracting the simulated neutron background, the purely exponential spectra reveal the contribution from a $\beta$-decay of $^{12}$B. Top panel (a) shows the detection rate (before correcting for the $\beta$-ray detection efficiency), while bottom panel (b) shows the decay rate (after correcting for the $\beta$-ray detection efficiency).}
\label{fig1}
\end{figure}

Figure~\ref{fig1} shows the experimental data for both detectors, before and after subtracting the neutron background. The top panel (a) shows the spectra uncorrected for the $\beta$-ray detection efficiency $\varepsilon_\beta$, in order to facilitate the visual separation of exponential fits. The bottom panel (b) shows the data corrected for the $\beta$-ray detection efficiency, emphasizing the consistency of results from two detectors. It is evident from fig.~\ref{fig1} that after the background subtraction the time distribution of the detected counts follows the expected exponential trend (with the expected lifetime), proving that no additional background components are present in the results.

According to the simulations, the contribution of $\beta$-rays from $^{12}$B produced outside the sample by the scattered neutrons is below 1\%. Though their contribution has been removed from the data through the neutron background subtraction, this source of uncertainty has been assigned a systematic value of 2\%. In addition, $^{13}$C content in natural carbon (with 1.1\% natural abundance) also contributes to the production of $^{12}$B and $^{13}$B (with decay properties very similar to those of $^{12}$B) through the (n,p), (n,d) and (n,np) reactions, whose cross sections are highly uncertain. No attempt was made to decouple their contribution from the measurements, except for the portion already included in the neutron background. Therefore, a 3\% systematic uncertainty related to the contribution of $^{13}$C has been assigned to the number of produced $^{12}$B nuclei. Finally, a 6\% systematic uncertainty was adopted in order to account for the uncertainty in both the simulated $\beta$-ray detection efficiency and the simulated neutron background. Combining the results from both detectors, the final value is $N_{^{12}\mathrm{B}}=68.5\pm0.4_\mathrm{stat}\pm4.8_\mathrm{syst}$.


\section{GEANT4 cross sections}
\label{geant}

It was already shown in Section \ref{analysis} that GEANT4 simulations play a central role in the experimental data analysis, in determining the weighting functions required by the PHWT, of identifying the neutron background and of determining the $\beta$-ray detection efficiency of the two C$_6$D$_6$ detectors. As will be shown in Section \ref{integral}, simulations are also indispensable in bringing the number of produced $^{12}$B nuclei into relation with the cross section of the $^{12}$C(\emph{n,p})$^{12}$B reaction and in exerting control over the same procedure. In that, the (effective) cross sections for $^{12}$C(n,p)$^{12}$B reaction will play an important role. (By effective we mean the cross sections which are not directly sampled from the preexisting data, but arise as an end-product of the model calculations.) Five different models from GEANT4 were considered in this work: HP (High Precision) package, Binary cascade, Bertini cascade, INCL++/ABLA model (INCL intranuclear cascade coupled to the ABLA deexcitation model) and QGS (Quark-Gluon-String) model \cite{manual}. GEANT4 version 9.6.p01 was used for all models except for INCL++/ABLA, which was run in GEANT4 version 10.0 (the only reason being that the INCL++/ABLA model is not available in version 9.6.p01). It should be remarked that the reliability of the total and elastic scattering cross sections adopted by GEANT4 is of crucial importance for the quality of subsequent analysis, directly affecting both the self-shielding and the multiple scattering factor discussed in Section~\ref{integral}.

\begin{figure}[b!]
\includegraphics[angle=90,width=1.0\linewidth,keepaspectratio]{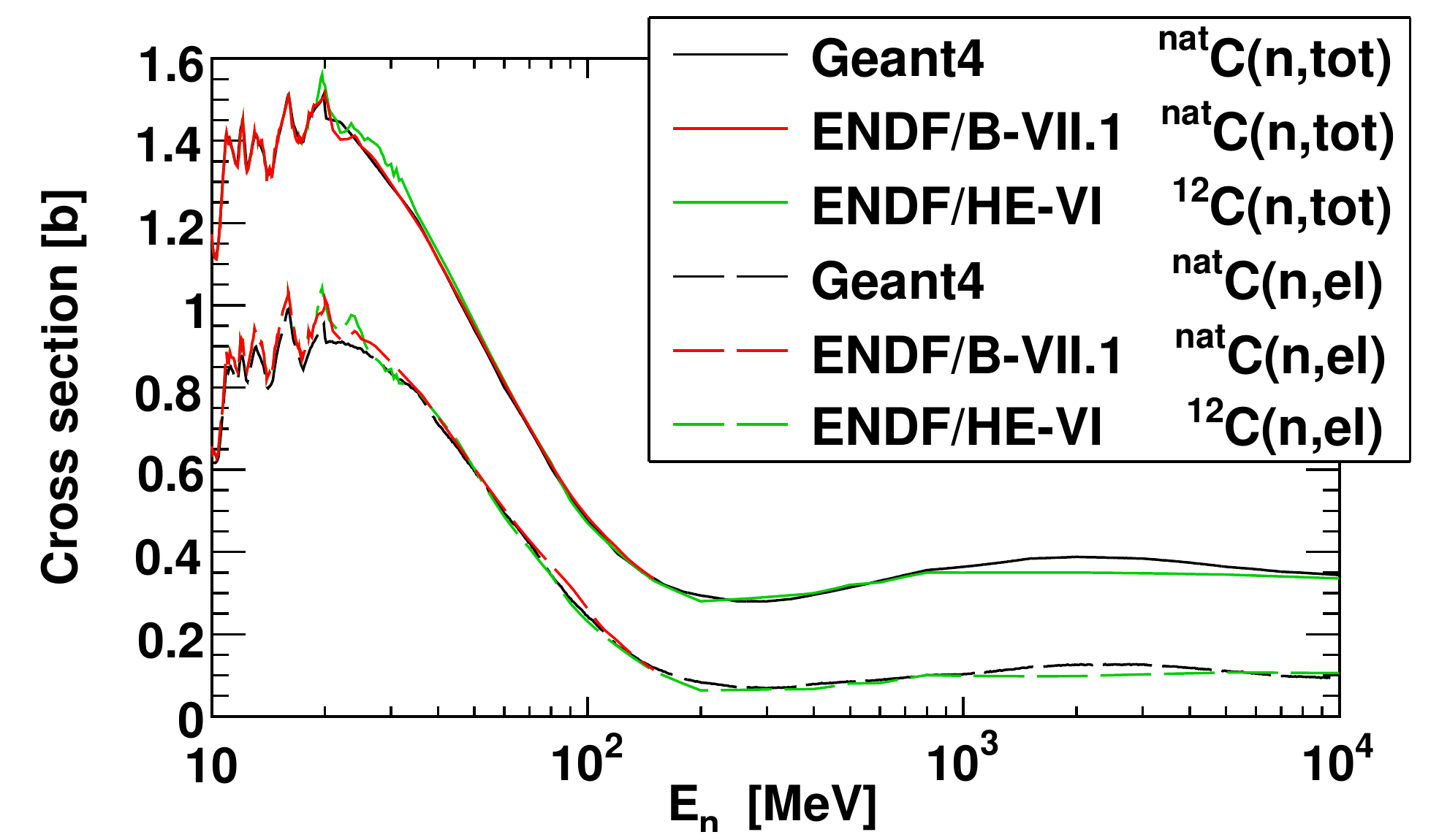}
\caption{Total cross section for the neutron induced reactions on $^\mathrm{nat}$C and the elastic scattering cross section extracted from GEANT4 simulations, compared to ENDF/B-VII.1 \cite{endf} and ENDF/HE-VI \cite{endfhe} libraries (data from ENDF/HE-VI are given for $^{12}$C).}
\label{fig3}
\end{figure}

The cross sections for all reactions of interest were extracted from GEANT4 by the following procedure. The $^\mathrm{nat}$C sample was irradiated in the simulation with neutrons following the exact energy dependence of the n\_TOF flux. The occurrence $N_r(E_\mathrm{n})$  of a given reaction $r$ was counted only if $r$ was the first reaction to take place. In parallel, the occurrence $N_\mathrm{tot}(E_\mathrm{n})$ of any reaction  in the sample was also counted. The probability for reaction $r$ without multiple scattering effects, \emph{i.e.} the first-chance yield $Y_r(E_\mathrm{n})$ at neutron energy $E_\mathrm{n}$ may be expressed as:
\begin{linenomath}\begin{equation}
\label{eq2}
Y_r(E_\mathrm{n})=(1-e^{-n\sigma_\mathrm{tot}(E_\mathrm{n})})\frac{\sigma_r(E_\mathrm{n})}{\sigma_\mathrm{tot}(E_\mathrm{n})}=\frac{N_r(E_\mathrm{n})}{N_0(E_\mathrm{n})}
\end{equation}\end{linenomath}
The first expression is given in terms of the reaction cross section $\sigma_r(E_\mathrm{n})$, total cross section $\sigma_\mathrm{tot}(E_\mathrm{n})$ and an areal density $n$ of the sample (in number of atoms per unit surface, which for the used $^\mathrm{nat}$C sample is equal to $n=0.114$~atoms/barn). The second part of eq.~(\ref{eq2}) expresses the probability as the ratio between the number $N_r(E_\mathrm{n})$ of times the reaction has occurred and the total number $N_0(E_\mathrm{n})$ of incident neutrons with energy $E_\mathrm{n}$. In order to be able to evaluate the particular cross section $\sigma_r(E_\mathrm{n})$, the total cross section $\sigma_\mathrm{tot}(E_\mathrm{n})$ must first be determined by inverting eq.~(\ref{eq2}), considering all neutron interactions in the sample:
\begin{linenomath}\begin{equation}
\label{eq3}
\sigma_\mathrm{tot}(E_\mathrm{n})=-\frac{1}{n}\ln\left(1-\frac{N_\mathrm{tot}(E_\mathrm{n})}{N_0(E_\mathrm{n})}\right)
\end{equation}\end{linenomath}
Ultimately, adopting $\sigma_\mathrm{tot}(E_\mathrm{n})$ from simulations (as opposed to any particular database) allows for the self-consistent calculation of $\sigma_r(E_\mathrm{n})$ of interest:
\begin{linenomath}\begin{equation}
\label{eq4}
\sigma_r(E_\mathrm{n})=\frac{N_r(E_\mathrm{n})}{N_\mathrm{tot}(E_\mathrm{n})}\sigma_\mathrm{tot}(E_\mathrm{n})
\end{equation}\end{linenomath}

Figure~\ref{fig3} compares the total and elastic cross section extracted from GEANT4 -- by eq.~(\ref{eq3}) and eq.~(\ref{eq4}), respectively -- with those from ENDF/B-VII.1 \cite{endf} and ENDF/HE-VI \cite{endfhe} library. It should be noted that the data in ENDF/B-VII.1 are available for $^\mathrm{nat}$C, while in ENDF/HE-VI for $^{12}$C. It is evident that the simulations closely reproduce the tabulated cross sections. Therefore, these results may be used with a high degree of confidence in identifying both the self-shielding and the multiple scattering factor described in the following Section.

\begin{figure}[t!]
\includegraphics[width=1.0\linewidth,keepaspectratio]{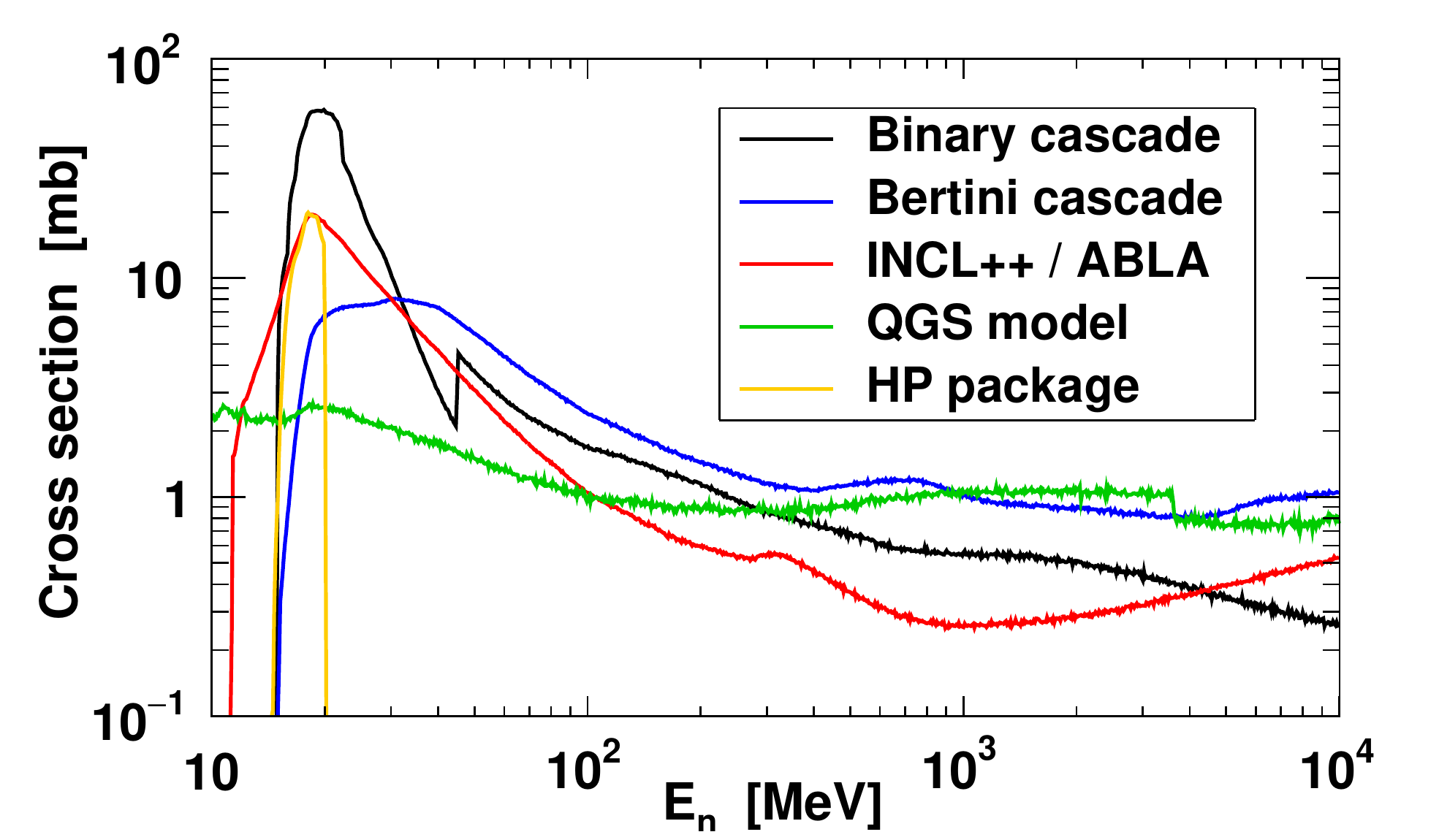}
\caption{Cross sections for $^\mathrm{nat}$C(n,p)$^{12}$B reaction from 5 different GEANT4 models (see text for details).}
\label{fig2}
\end{figure}

The (effective) cross sections for the $^{12}$C(n,p)$^{12}$B reaction have also been calculated on the basis of eq.~(\ref{eq4}). They are shown in fig.~\ref{fig2}. In particular, the reconstructed cross section from the HP package builds confidence that the procedure was correctly performed, since the HP package adopts the cross sections from the ENDF/B-VII.1 database \cite{endf} (see fig.~\ref{fig0}). It should be noted that the cross sections for the $^{12}$C(n,p)$^{12}$B reaction extracted from GEANT4 are not of particular interest to this work, and in some cases are clearly unreliable. For example, this is the case of the one extracted from the quark-gluon-string models, which clearly cannot be applied below several GeV. Nevertheless, all of them extend over the whole energy range from the reaction threshold up to 10~GeV, which makes them useful in determination of the multiple scattering correction discussed in Section~\ref{integral}, and in illustrating the robustness of these results against the wildly varying cross sections.

\section{Comparison with models and evaluations}
\label{integral}

The produced number of $^{12}$B nuclei per neutron bunch is related to the $^{12}$C(n,p)$^{12}$B reaction cross section through the following expression:
\begin{linenomath}\begin{equation}
\label{eq5}
N_{{^{12}\mathrm{B}}}=\int_{13.6\:\mathrm{MeV}}^{{10\:\mathrm{GeV}}}\frac{1-e^{-n\sigma_\mathrm{tot}(E_\mathrm{n})}}{\sigma_\mathrm{tot}(E_\mathrm{n})}\eta(E_\mathrm{n})\phi(E_\mathrm{n})\sigma(E_\mathrm{n})\mathrm{d}E_\mathrm{n}
\end{equation}\end{linenomath}
where the energy range covered by the integral spans from the reaction threshold at 13.6~MeV up to the highest neutron energy provided by the n\_TOF beam, \emph{i.e.} 10~GeV. The product of the first term (consisting of the self-shielding factor divided by the total cross section $\sigma_\mathrm{tot}(E_\mathrm{n})$; $n$ being the areal density of sample in number of atoms per unit surface) with the cross section $\sigma(E_\mathrm{n})$ represents the first-chance reaction yield per incident neutron of energy $E_\mathrm{n}$, which does not take into account the multiple scattering effect. This is accounted for separately, by the multiple scattering factor $\eta(E_\mathrm{n})$ which depends mostly on the elastic cross section. Finally, the sample-incident neutron flux $\phi(E_\mathrm{n})$ also has to be considered.


\subsection{Self-shielding factor}

The self-shielding factor, appearing as the numerator from eq.~\ref{eq5}, is central in defining the first-chance yield of the $^{12}$C(n,p)$^{12}$B reaction. It determines the relative portion of the neutron beam that is attenuated after passing through the full length of the sample. Equivalently, it may be considered as a probability for a single neutron to initiate any possible reaction in the sample -- as reflected through the adoption of the total cross section $\sigma_\mathrm{tot}(E_\mathrm{n})$ -- and be removed from the incident neutron beam. On the other hand, the ratio $\sigma(E_\mathrm{n})/\sigma_\mathrm{tot}(E_\mathrm{n})$ represents the probability that among all possible reactions, the particular one with the cross section $\sigma(E_\mathrm{n})$ is to take place. The reliability of the total cross section adopted in GEANT4 has already been confirmed in Section~\ref{geant} (see fig.~\ref{fig3}).

\subsection{Neutron flux}
\label{flux}

\begin{figure}[b!]
\includegraphics[width=1.0\linewidth,keepaspectratio]{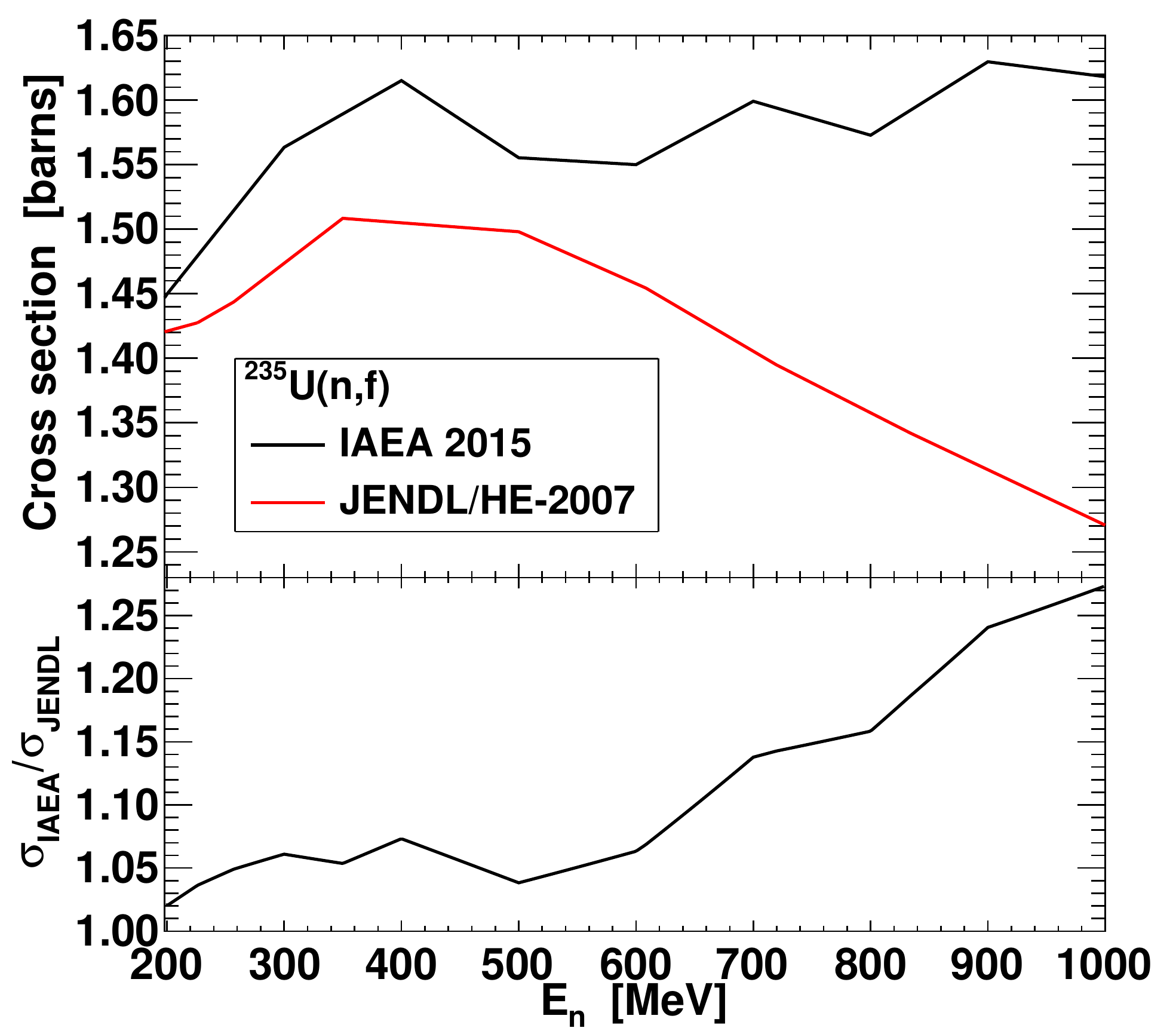}
\caption{Top panel: $^{235}$U(n,f) cross section from JENDL/HE-2007 compared with the recently extended reference cross section from IAEA. Bottom panel: ratio between cross sections from the top panel.}
\label{fig10}
\end{figure}

\begin{figure}[t!]
\includegraphics[width=1.0\linewidth,keepaspectratio]{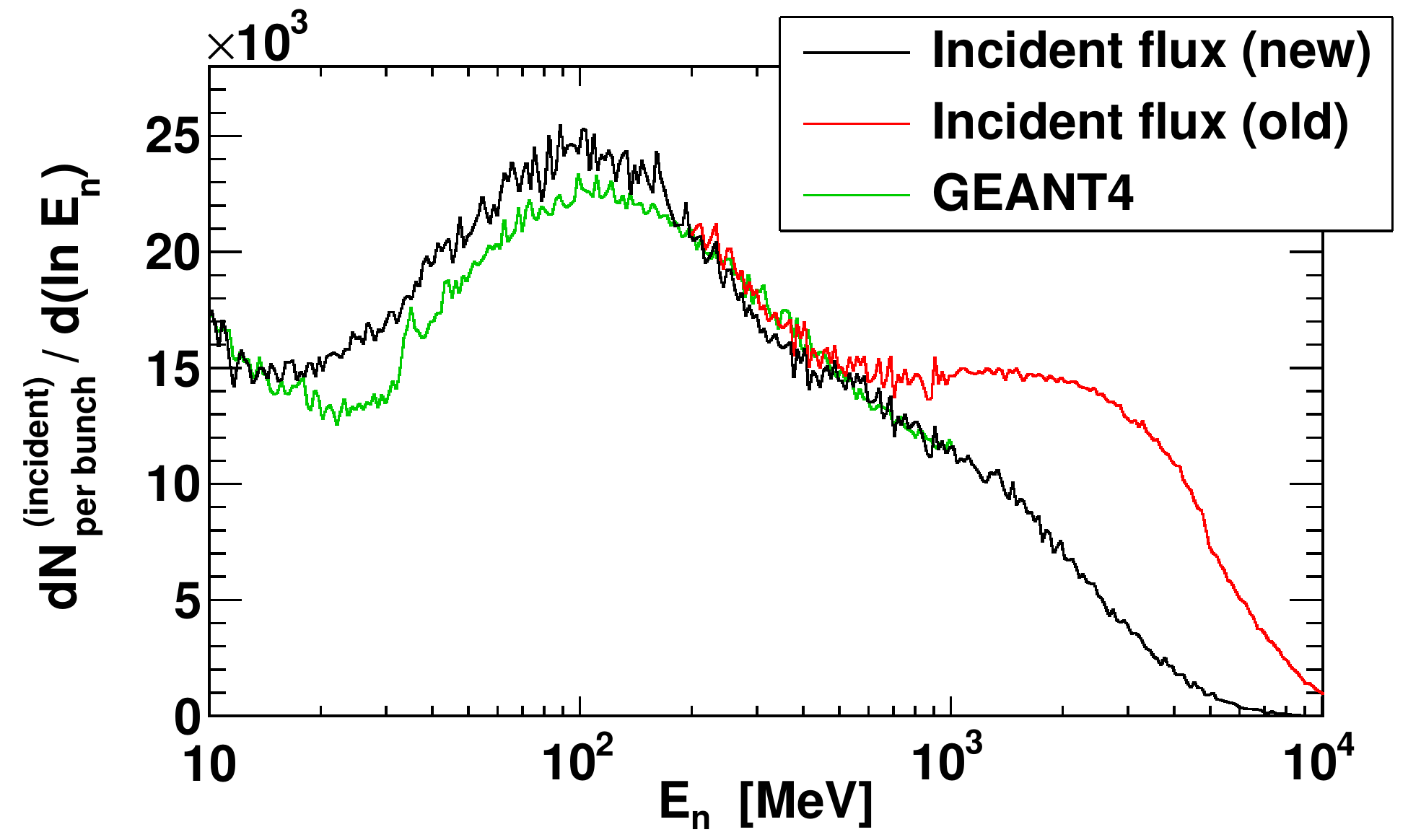}
\caption{Sample-incident neutron flux between 10~MeV and 10~GeV. The total flux up to 1~GeV was measured using PPAC detector and normalized to the $^{235}$U(n,f) yield calculated from the associated cross section. The flux above 200~MeV was reevaluated, based on the recently extended $^{235}$U(n,f) reference cross section from IAEA. In the energy range above 1~GeV the normalized results from GEANT4 simulations are used. See the main text for the details.}
\label{fig4}
\end{figure}

The neutron flux $\phi(E_\mathrm{n})$ entering the experimental area was measured up to 1~GeV by means of Parallel Plate Avalanche Counters \cite{ppac1,ppac2}, relying on the $^{235}$U(n,f) reaction. In Refs.~\cite{carlos,massimo} the n\_TOF flux was determined on the basis of the standard cross section from the ENDF/B-VII.1 evaluation \cite{endf} up to 200~MeV. Above this energy, and up to~1 GeV, the JENDL/HE-2007 database \cite{jendl} was used. Recently, a new evaluation has been proposed by  IAEA as a reference cross section \cite{standard}, now widely considered to be more reliable. Therefore, it was decided to re-evaluate the experimental flux above 200~MeV using the new IAEA cross section reference. Figure~\ref{fig10} compares the cross sections from JENDL/HE-2007 library, used in the past, with the newly extended reference from IAEA. The ratio between them is also displayed in the figure. Previously evaluated flux simply had to be divided by this ratio in order to obtain the reevaluated flux. Figure~\ref{fig4} compares the new flux (incident on the sample) with the one used in the past. The beam interception factor, required for identifying the portion of the flux incident on the sample, was determined as a function of neutron energy from FLUKA~\cite{fluka} simulations of the neutron transport after the spallation process, and adjusted to the experimental data at low energies using the saturated resonance technique \cite{au_bif} applied to the 4.9~eV neutron capture resonance of $^{197}$Au. For clarity, the neutron flux from fig.~\ref{fig4} is shown in units of lethargy, from which the flux per unit energy may be obtained as:
\begin{linenomath}\begin{equation}
\phi(E_\mathrm{n})=\frac{dN_\mathrm{per\:bunch}}{dE_\mathrm{n}}=\frac{1}{E_\mathrm{n}}\times \frac{dN_\mathrm{per\:bunch}}{d\ln E_\mathrm{n}}
\end{equation}\end{linenomath}
where $N_\mathrm{per\:bunch}$ is the number of neutrons per neutron bunch, reaching the experimental area (in case of the evaluated flux) or impinging on the sample (in case of the incident flux).

While the flux was measured up to 1~GeV, for the purpose of this work it is necessary to consider the full energy distribution of the n\_TOF beam, since all neutrons above 13.6~MeV contribute to the $^{12}$C(n,p)$^{12}$B reaction and the production of $^{12}$B nuclei. The neutron flux above 1~GeV can be estimated by means of dedicated simulations of the spallation process. It was found that the results from recently developed GEANT4 simulations of the spallation process \cite{submitted} are in good agreement with the shape of the flux around 1~GeV. For this reason, the GEANT4 simulations have been adopted for extending the evaluated flux up to 10~GeV. In ref.~\cite{submitted} the results from several different physics lists have been compared, namely from \mbox{FTFP\_BERT\_HP}, \mbox{FTFP\_INCLXX\_HP}, \mbox{QGSP\_BERT\_HP}, \mbox{QGSP\_BIC\_HP} and \mbox{QGSP\_INCLXX\_HP}. The \mbox{QGSP\_INCLXX\_HP} list was found to better reproduce the absolute scale of the neutron flux, within the overall energy range from thermal up to 10~GeV. However, the \mbox{FTFP\_BERT\_HP} list provides the best reproduction of the shape of the experimental flux at high energies. For this reason, the \mbox{FTFP\_BERT\_HP} physics list has been used in this work in order to extend the evaluated flux beyond 1~GeV. These results have been normalized so as to match the experimental flux at 1~GeV. The normalized flux is also shown in fig.~\ref{fig4}. It was also multiplied by the energy dependent beam interception factor, in order to translate it into the sample incident flux.

\subsection{Multiple scattering factor}
\label{multiple}

\begin{figure}[b!]
\vspace*{7mm}
\begin{overpic}[width=1.0\linewidth,keepaspectratio]{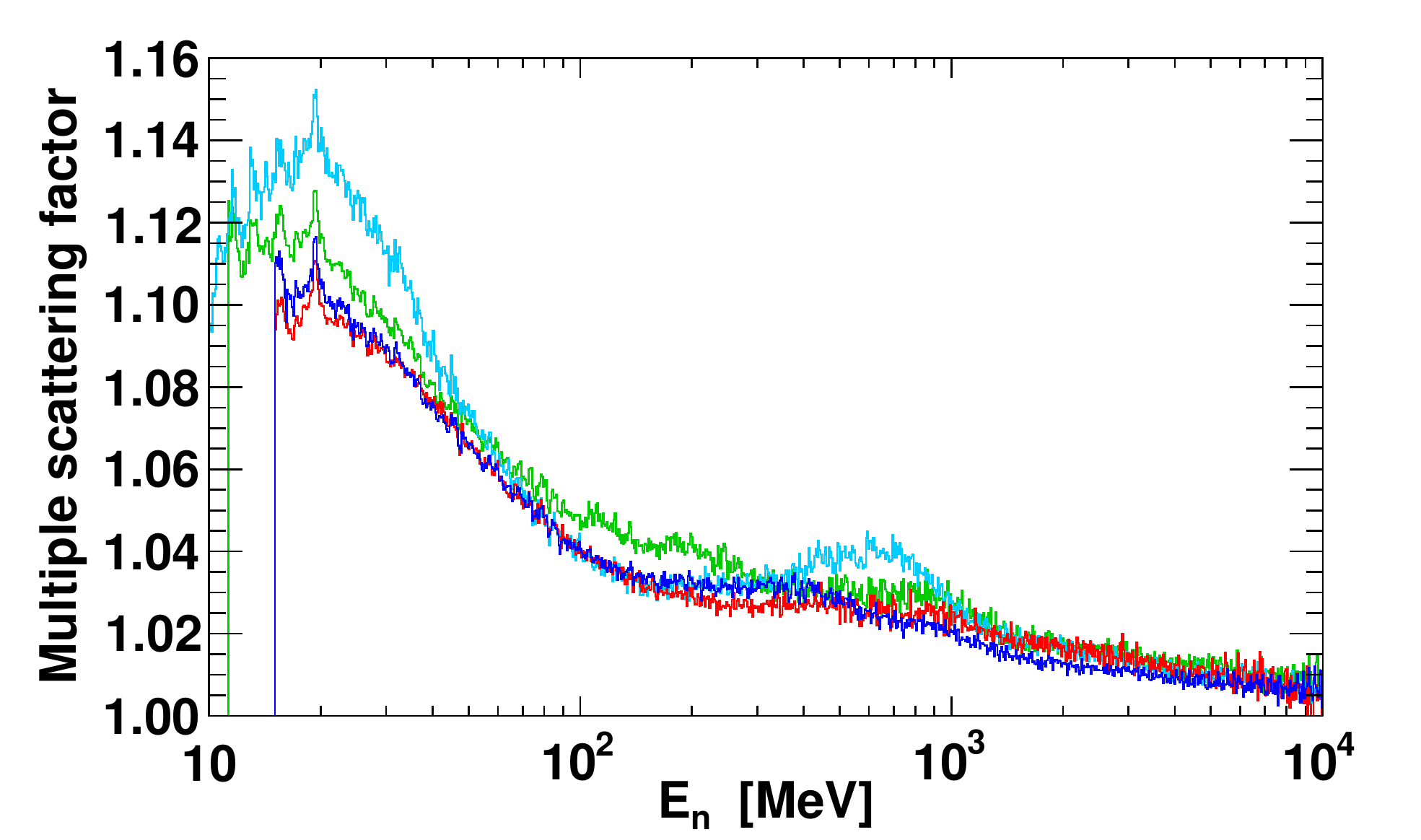}
\put(0,0){\includegraphics[width=1.0\linewidth,keepaspectratio]{multiple.pdf}} 
\put(36.25,30.1){\includegraphics[width=0.62\linewidth,keepaspectratio]{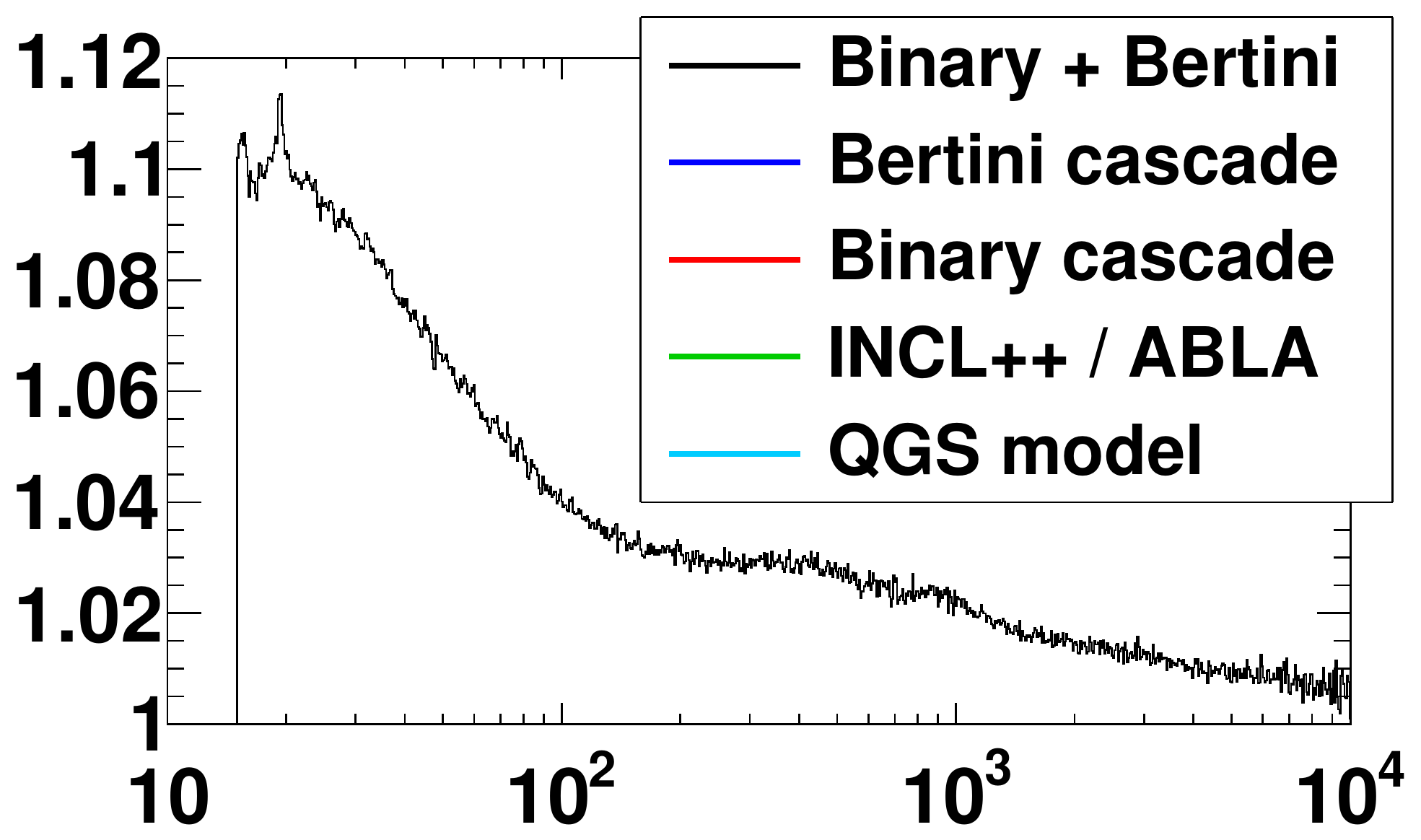}}
\end{overpic}
\caption{Multiple scattering factor from four different GEANT4 models. Inset shows the final dependence adopted for calculations, obtained by taking the average between the results from Binary cascade and Bertini cascade.}
\label{fig5}
\end{figure}

The multiple scattering factor $\eta(E_\mathrm{n})$ was determined from GEANT4 simulations. We remark that in the simulations it is of crucial importance to use a realistic flux, \emph{i.e.} the one shown in fig.~\ref{fig4}, since -- after the (in)elastic scattering -- the neutrons of initially higher energy are feeding the lower energy portions of the spectrum. In the simulations the primary neutron energy $E_0$ was recorded, together with the final energy $E_\mathrm{n}$ of the neutron giving rise to  the $^\mathrm{nat}$C(n,p)$^{12}$B reaction. The multiple scattering factor -- increasing the $^{12}$B production at the neutron energy $E_\mathrm{n}$ -- was determined as the ratio between the number $N(E_\mathrm{n})$ of $^{12}$B nuclei produced by all neutrons of energy $E_\mathrm{n}$ and the number $N(E_\mathrm{n}=E_0)$ of those produced by neutrons not previously affected by scattering (\emph{i.e.} those with the final energy equal to the primary one):
\begin{linenomath}\begin{equation}
\eta(E_\mathrm{n})=\frac{N(E_\mathrm{n})}{N(E_\mathrm{n}=E_0)}
\end{equation}\end{linenomath}
In order to investigate the stability and reliability of results, the simulations were performed using 4 different inelastic scattering models that extend over the full energy range of interest -- Binary cascade, Bertini cascade, INCL++/ABLA model and QGS model \cite{manual}. Model predictions are compared in fig.~\ref{fig5}. It is evident that the models yielding substantially different effective cross sections (see fig.~\ref{fig2}) give very consistent multiple scattering corrections up to 11\% and 15\%, depending in the model used. This is due to the fact that the multiple scattering factor is dominantly affected by the elastic scattering. The reliability of the elastic scattering cross section from GEANT4 has already been confirmed in Section~\ref{geant} (see fig.~\ref{fig3}). Considering the consistency between the multiple scattering factors from Bertini cascade and Binary cascade, their average value was adopted -- shown in the inset of fig.~\ref{fig5} -- that yields a maximal multiple scattering correction of approximately 11\%, peaking around 20~MeV. Additional reason for using the combination of these two cascade models is that their combined result will later be shown to best describe the experimental n\_TOF data.

\subsection{Weighting function}

\begin{figure}[b!]
\includegraphics[width=1.0\linewidth,keepaspectratio]{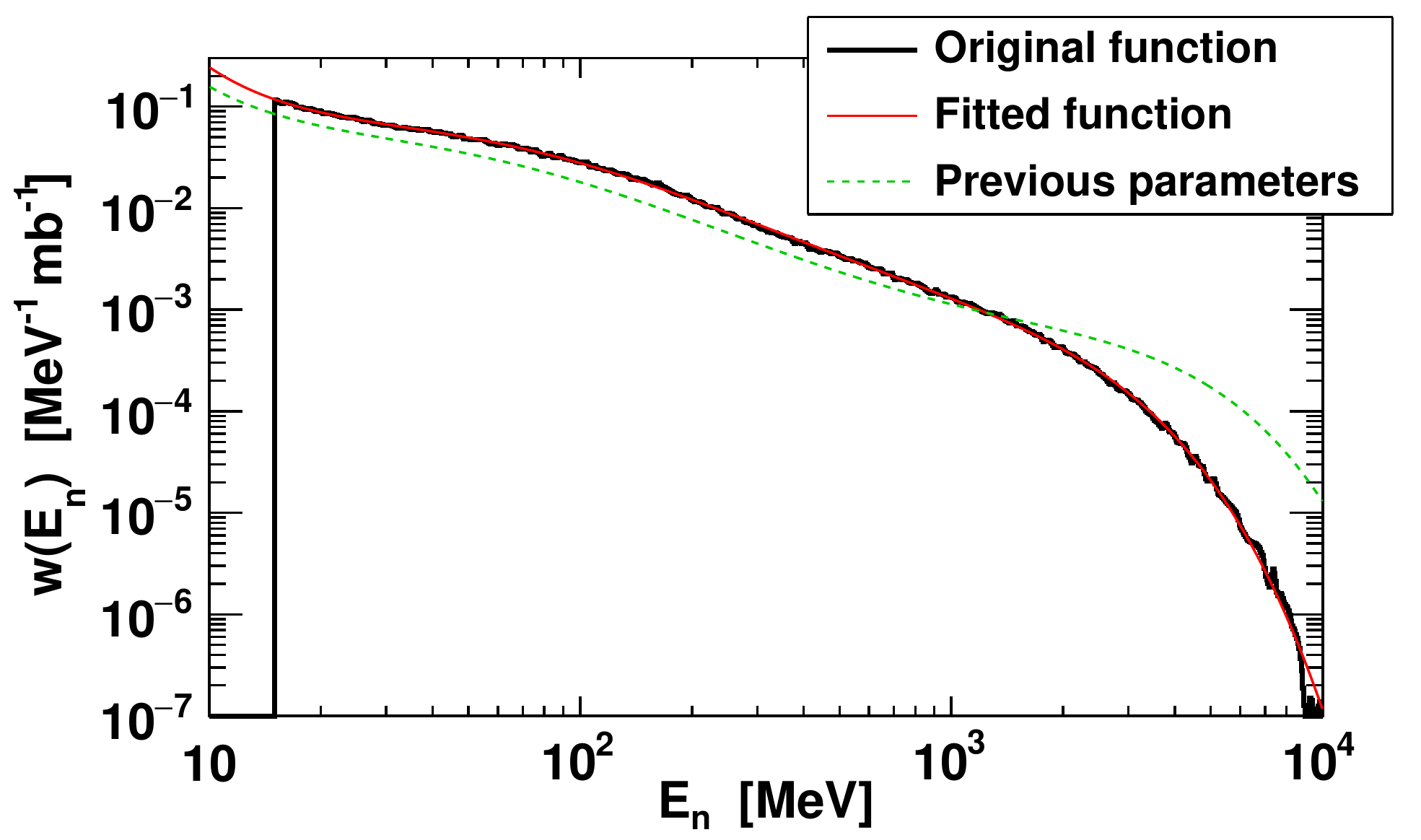}
\caption{Weighting function used for calculating the number of produced $^{12}$B nuclei from the cross section for the $^{12}$C(n,p)$^{12}$B reaction. Fitted function is the fifth degree polynomial. The dashed line shows the function reproduced from the less precise set of parameters reported in ref.~\cite{nprc}.}
\label{fig6}
\end{figure}

The product of all terms in eq. (\ref{eq5}) -- apart from the cross section $\sigma(E_\mathrm{n})$ -- may be treated as a unique weighting function $w(E_\mathrm{n})$ (which is not to be confused with the weighting function from the PHWT):
\begin{linenomath}\begin{equation}
\label{eq7}
w(E_\mathrm{n})=\frac{1-e^{-n\sigma_\mathrm{tot}(E_\mathrm{n})}}{\sigma_\mathrm{tot}(E_\mathrm{n})}\eta(E_\mathrm{n})\phi(E_\mathrm{n})
\end{equation}\end{linenomath}
This function was fitted to the fifth degree polynomial:
\begin{linenomath}\begin{equation}
\label{eq6}
\log_{10}\frac{w(E_\mathrm{n})}{w_0}=\sum_{m=0}^5a_m\left(\log_{10}\frac{E_\mathrm{n}}{E_0}\right)^m
\end{equation}\end{linenomath}
with $E_0=1$~MeV and $w_0=1$~MeV$^{-1}$ mb$^{-1}$. The fit parameters have already been reported in ref.~\cite{nprc}. However, an oversight was made therein, which consists in reporting only 3 most significant digits. It was later found that this was not a sufficient level of precision for a successful reconstruction of the original weighting function and that a minimum of 4 digits had to be used. In addition to reevaluating the neutron flux, since reporting the results in ref.~\cite{nprc}, slight improvements were made to the simulation of the multiple scattering factor (the latest results being shown in fig.~\ref{fig5}), leading to a change in the weighting function. Globally, this change amounts to approximately 1\%. The latest weighting function $w(E_\mathrm{n})$ is shown in fig.~\ref{fig6}, together with the correct fit and the one produced by the previous, less precise set of parameters from ref.~\cite{nprc}. The parameters from both the latest polynomial fit and the one from ref.~\cite{nprc} are now reported to a greater degree of precision in table~\ref{tab1}. The weighting function was assigned 5\% systematic uncertainty due to the contribution from the self-shielding and the multiple scattering factor, and an additional 6\% uncertainty due to the neutron flux, resulting in an 8\% total systematic uncertainty.

\begin{table}[t!]
\caption{Parameters of the polynomial fit from eq.~(\ref{eq6}). The previous values refer to the parameters reported in ref.~\cite{nprc}.}
\label{tab1}
\centering
\begin{tabular}{ccc}
\hline\hline
Parameter&Previous value&Latest value\\
\hline
$a_0$&10.2225&12.9676\\
$a_1$&--27.4508&--33.9199\\
$a_2$&26.3467&32.3332\\
$a_3$&--12.3142&--15.0657\\
$a_4$&2.72984&3.36573\\
$a_5$&--0.232123&--0.291966\\
\hline\hline
\end{tabular}
\end{table}

\begin{figure}[b!]
\includegraphics[width=1.0\linewidth,keepaspectratio]{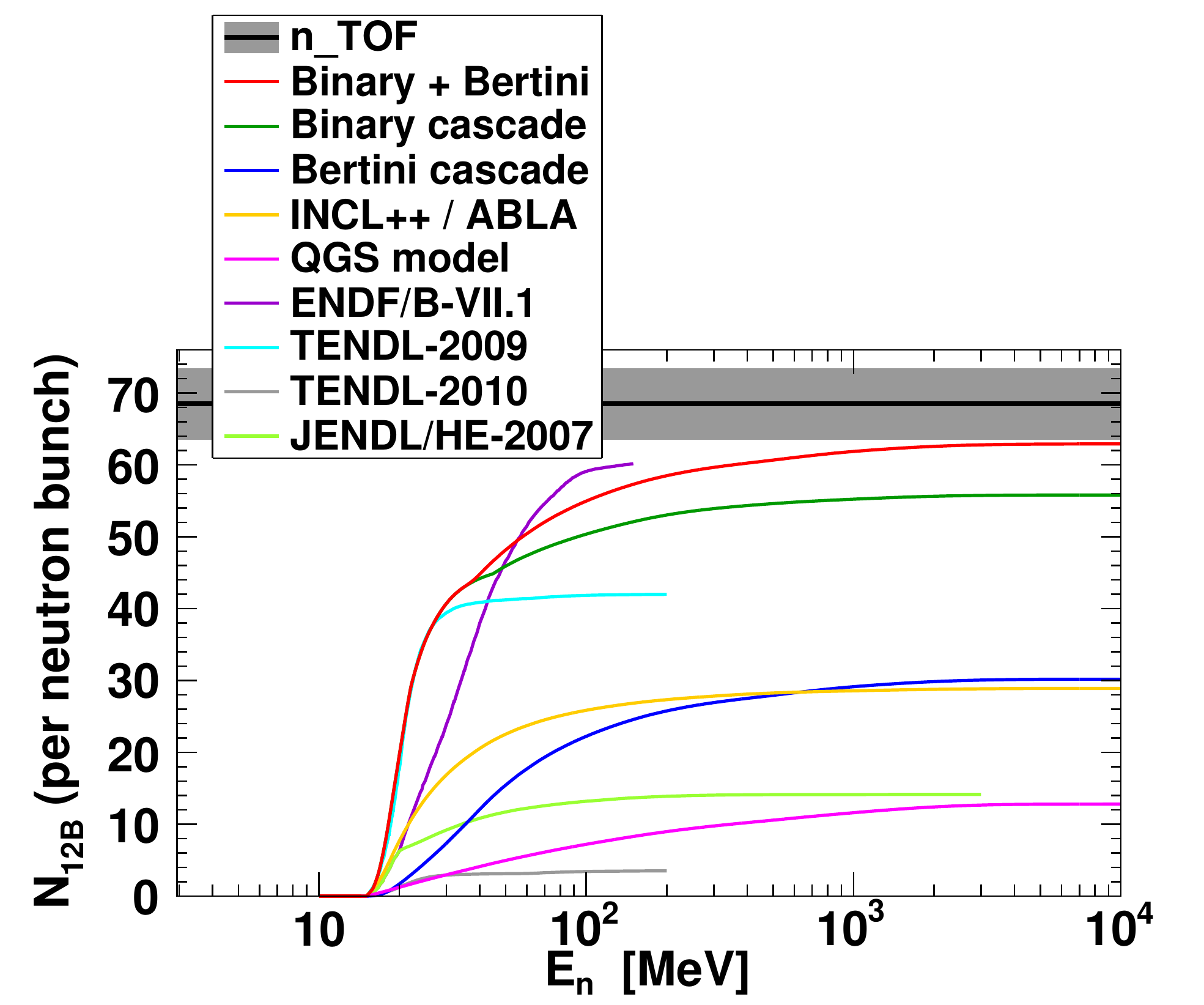}
\caption{Cumulative number of $^{12}$B nuclei produced by different GEANT4 models and those calculated from several evaluation libraries. The final values at 10~GeV are to be compared with the experimental n\_TOF result (the shaded area shows its standard uncertainty range). All cumulative distributions are assigned a systematic uncertainty of 8\%, inherited from the weighting function $w(E_\mathrm{n})$. The curve coming closest to the n\_TOF datum corresponds to the combination of Binary cascade below 30~MeV and Bertini cascade above 30~MeV.}
\label{fig7}
\end{figure}

Having obtained the weighting function, the number of produced $^{12}$B nuclei may be calculated for every GEANT4 model, starting from the known cross sections shown in fig.~\ref{fig2}. Furthermore, running the simulations (using any of the models) with full geometric setup -- \emph{i.e.} irradiating the sample and subsequently detecting the $\beta$-rays by two C$_6$D$_6$ detectors -- one may perform exactly the same analysis over the simulated data as was performed over the experimental data. Therefore, for the same model the number of produced $^{12}$B nuclei may be extracted both by fitting the exponential decay spectra and by a direct integration of eq.~(\ref{eq5}), since the model cross section is known in advance. This allowed to cross check the calculation of the weighting function from eq.~(\ref{eq7}), since any error would lead to inconsistent results. The combination of models that best reproduces the experimental result is the combination of Binary cascade below 30~MeV and Bertini cascade above 30~MeV (a detailed list of the neutron physics models used in this work may be found in ref.~\cite{background}; we advise the reader that QBBC is in this sense the closest of the prearranged physics lists available in GEANT4, which makes the smooth transition between the Binary cascade and Bertini cascade around 1~GeV). All other currently available models underestimate the number of produced $^{12}$B nuclei \cite{nprc}, in particular at low energy, while this particular combination maximizes it, as it can be also noticed from the cross section in fig.~\ref{fig2}. Figure~\ref{fig7} also demonstrates this, by showing the cumulative number of produced $^{12}$B nuclei:
\begin{linenomath}\begin{equation}
\label{eq11}
N_{{^{12}\mathrm{B}}}^{(\mathcal{E}_n)}=\int_{13.6\:\mathrm{MeV}}^{\mathcal{E}_n}w(E_\mathrm{n})\sigma(E_\mathrm{n})\mathrm{d}E_\mathrm{n}
\end{equation}\end{linenomath}
for different models and evaluation libraries. The final values $N_{^{12}\mathrm{B}}^{(10\:\mathrm{GeV})}$ for different GEANT4 models are given in table~\ref{tab2}.

\subsection{Resonance integral}
\label{resonance}

Another value that may be extracted from the experimental n\_TOF result is the quantity $I_{{^{12}\mathrm{B}}}$ analogous to the resonance integral:
\begin{linenomath}\begin{equation}
\label{eq10}
I_{{^{12}\mathrm{B}}}=\int_{13.6\:\mathrm{MeV}}^{{10\:\mathrm{GeV}}}\frac{\sigma(E_\mathrm{n})}{E_\mathrm{n}}\mathrm{d}E_\mathrm{n}
\end{equation}\end{linenomath}
which is widely used in the nuclear reactor physics. It is related to the number of produced $^{12}$B nuclei in a manner:
\begin{linenomath}\begin{equation}
\label{eq8}
N_{{^{12}\mathrm{B}}}\approx \kappa\times I_{{^{12}\mathrm{B}}}
\end{equation}\end{linenomath}
where $\kappa$ is the conversion factor. Evidently, the resonance integral approach simplifies the comparison of the model predictions of the $^{12}$C(n,p)$^{12}$B cross section with the integrated experimental result. The conversion factor may be estimated from the weighting function itself, as:
\begin{linenomath}\begin{equation}
\label{eq9}
\kappa\approx\frac{\int_{13.6\:\mathrm{MeV}}^{{10\:\mathrm{GeV}}} w(E_\mathrm{n})\mathrm{d}E_\mathrm{n}}{\int_{13.6\:\mathrm{MeV}}^{{10\:\mathrm{GeV}}}\frac{\mathrm{d}E_\mathrm{n}}{E_\mathrm{n}}}
\end{equation}\end{linenomath}
This procedure yields the value of $\kappa=1.565$~mb$^{-1}$. However, since this is just the first approximation, a different method was used for finding $\kappa$, which relies on simulations and allows to examine the stability and robustness of the result against the varying cross sections for the $^{12}$C(n,p)$^{12}$B reaction. Starting from the known cross sections for 4 different GEANT4 models extending throughout the entire energy range between 13.6~MeV and 10~GeV (fig.~\ref{fig2}), the number $N_{^{12}\mathrm{B}}$ of produced $^{12}$B nuclei was calculated for every model, according to eq.~(\ref{eq5}). The resonance integral analogy from eq.~(\ref{eq8}) was also calculated from every model's cross section, yielding -- in ratio with $N_{^{12}\mathrm{B}}$ -- the value of $\kappa$ for each model separately. The combination of Binary cascade and Bertini cascade was also included in calculations, due to the integral result being closest to the experimental value. For all models, the number of produced $^{12}$B nuclei, the resonance integral analogy values and the conversion factors are listed in table~\ref{tab2}. All these values are assigned a relative uncertainty of 8\%, inherited from the systematic uncertainty in the weighting function $w(E_\mathrm{n})$. The result from a combined Binary/Bertini model was adopted as the final one, due to the closest agreement with the experimental value for the number of produced $^{12}$B nuclei. The spread of results from all considered models was used to determine the uncertainty in the final conversion factor: \mbox{$\kappa=1.86\pm0.1$~mb$^{-1}$}. From here follows the value of $I_{{^{12}\mathrm{B}}}=37\pm3$~mb for the resonance integral analogy calculated from the experimental n\_TOF data.

\begin{table}[t!]
\caption{Number $N_{^{12}\mathrm{B}}$ of $^{12}$B nulcei produced, the resonance integral analogy $I_{{^{12}\mathrm{B}}}$ and the transition factor \mbox{$\kappa=N_{^{12}\mathrm{B}}/I_{{^{12}\mathrm{B}}}$} from different GEANT4 models. All GEANT4 results are assigned a relative uncertainty of 8\%, inherited from the systematic uncertainty in the weighting function $w(E_\mathrm{n})$. Final results from the analysis of the experimental n\_TOF data are also listed.}
\label{tab2}
\centering
\begin{tabular}{cccc}
\hline\hline
Source&$N_{^{12}\mathrm{B}}$&$\kappa$ [mb$^{-1}$]&$I_{{^{12}\mathrm{B}}}$ [mb]\\
\hline
Binary cascade&55.82		&1.8671	&29.90\\
Bertini cascade&30.19		&1.9512	&15.47\\
INCL++/ABLA&28.90		&1.8302	&15.79\\
QGS model&12.81		&1.6250	&7.88\\
Binary/Bertini&62.92		&1.8579	&33.87\\
\hline
n\_TOF&68.5$\pm$4.8&1.86$\pm$0.1&37$\pm$3\\
\hline\hline
\end{tabular}
\end{table}

\section{TALYS-1.6 calculations}
\label{theory}

Since the cross sections of the reactions that can produce $^{12}$B nuclei in the $^\mathrm{nat}$C sample -- \emph{i.e.} $^{12}$C(n,p), $^{13}$C(n,np) and $^{13}$C(n,d) -- have their maxima in the energy range from 20 MeV to 40 MeV, it is not surprising that the $^{12}$B production cannot be accurately described by a quark-gluon string model, which is expected to work at incident energies larger than a few GeV. Furthermore, whereas at energies above $\sim$100~MeV the preequilibrium stage follows the one described by intranuclear cascade models, at energies below $\sim$100~MeV the reaction proceeds directly through the preequilibrium models implemented in Monte Carlo codes, which are by necessity (for performance reasons) somewhat simplistic, when compared to the fully quantum-mechanical calculations. Thus, we consider it worthwhile to carry out cross section calculations with the fully quantum-mechanical models contained in the TALYS-1.6 \cite{Ko12,Ko13} code, from 13.6 MeV -- the threshold energy of the $^{12}$C(n,p) reaction -- up to 200 MeV, where intranuclear cascade models are already considered to be valid.

\begin{figure}[b!]
\includegraphics[width=1.0\linewidth]{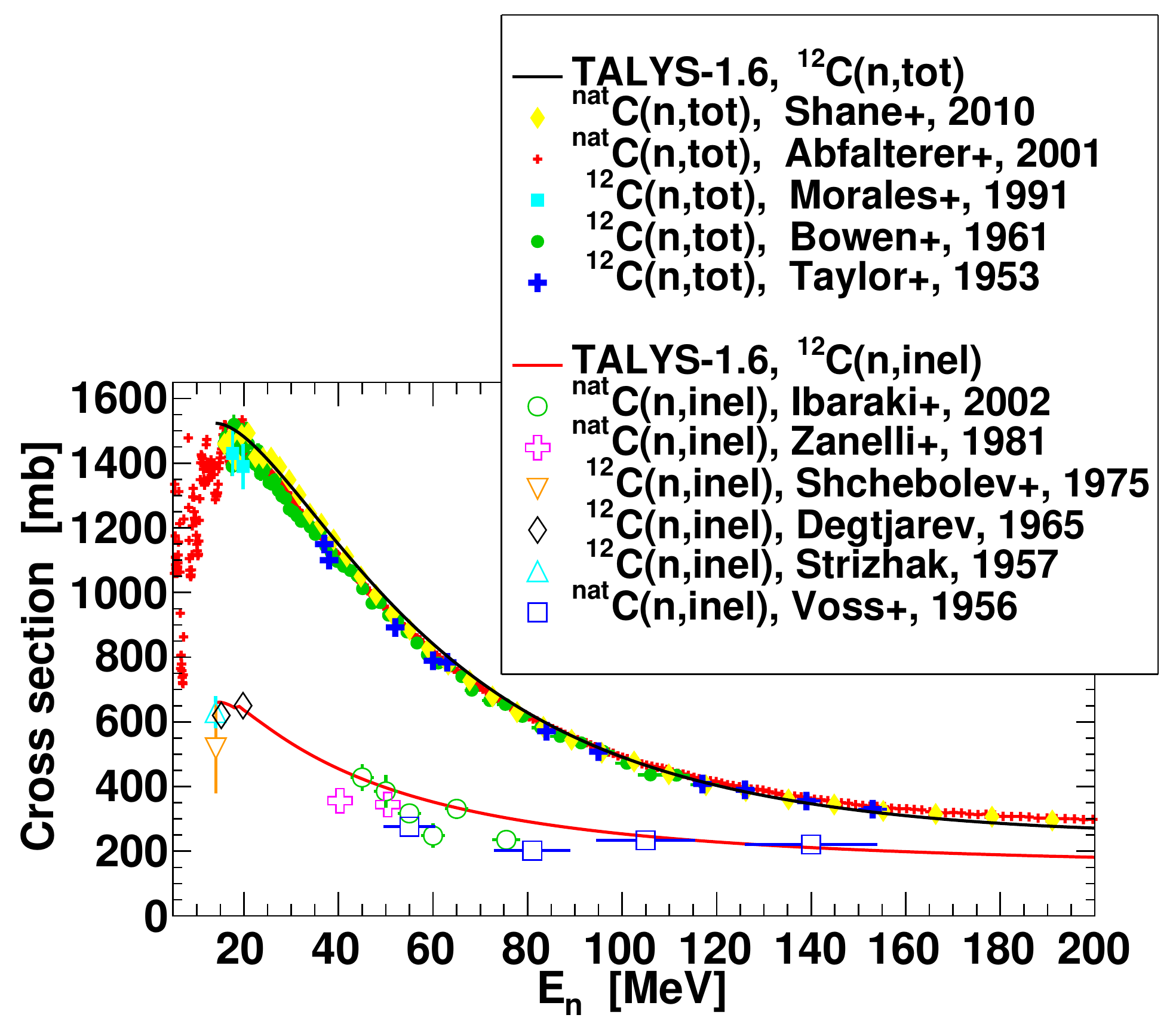}
\caption{$^{12}$C(n,total) and $^{12}$C(n,total inelastic) cross sections computed with TALYS-1.6, in comparison with experimental data for $^{12}$C and $^\mathrm{nat}$C.}
\label{fig8}
\end{figure}

Total, elastic and inelastic cross sections have been computed by means of an optical potential whose quality is shown in fig.~\ref{fig8}, where the total and total inelastic cross sections are compared with selected experimental data for both $^{12}$C and $^\mathrm{nat}$C. The theoretical total cross section is in excellent agreement with measurements on $^{12}$C -- such as those from Taylor and Wood \cite{Ta53}, Bowen \emph{et al.} \cite{Bo61}, Morales \emph{et al.} \cite{Mo91} -- and also with recent measurements on $^\mathrm{nat}$C, extended over the whole energy range of interest, among which are those from Shane \emph{et al.} \cite{Sh10} and Abfalterer \emph{et al.} \cite{Ab01}. In addition, fig.~\ref{fig8} shows that the theoretical inelastic cross section also compares well with the $^{12}$C measurements from Strizhak \cite{St57}, Degtjarev \cite{De65} and Shchebolev \emph{et al.} \cite{Sh75}, together with the $^\mathrm{nat}$C measurements in a larger energy range, from Voss and Wilson \cite{Vo56}, Zanelli \emph{et al.} \cite{Za81} and Ibaraki \emph{et al.} \cite{Ib02}.

\begin{figure}[b!]
\includegraphics[width=1.0\linewidth]{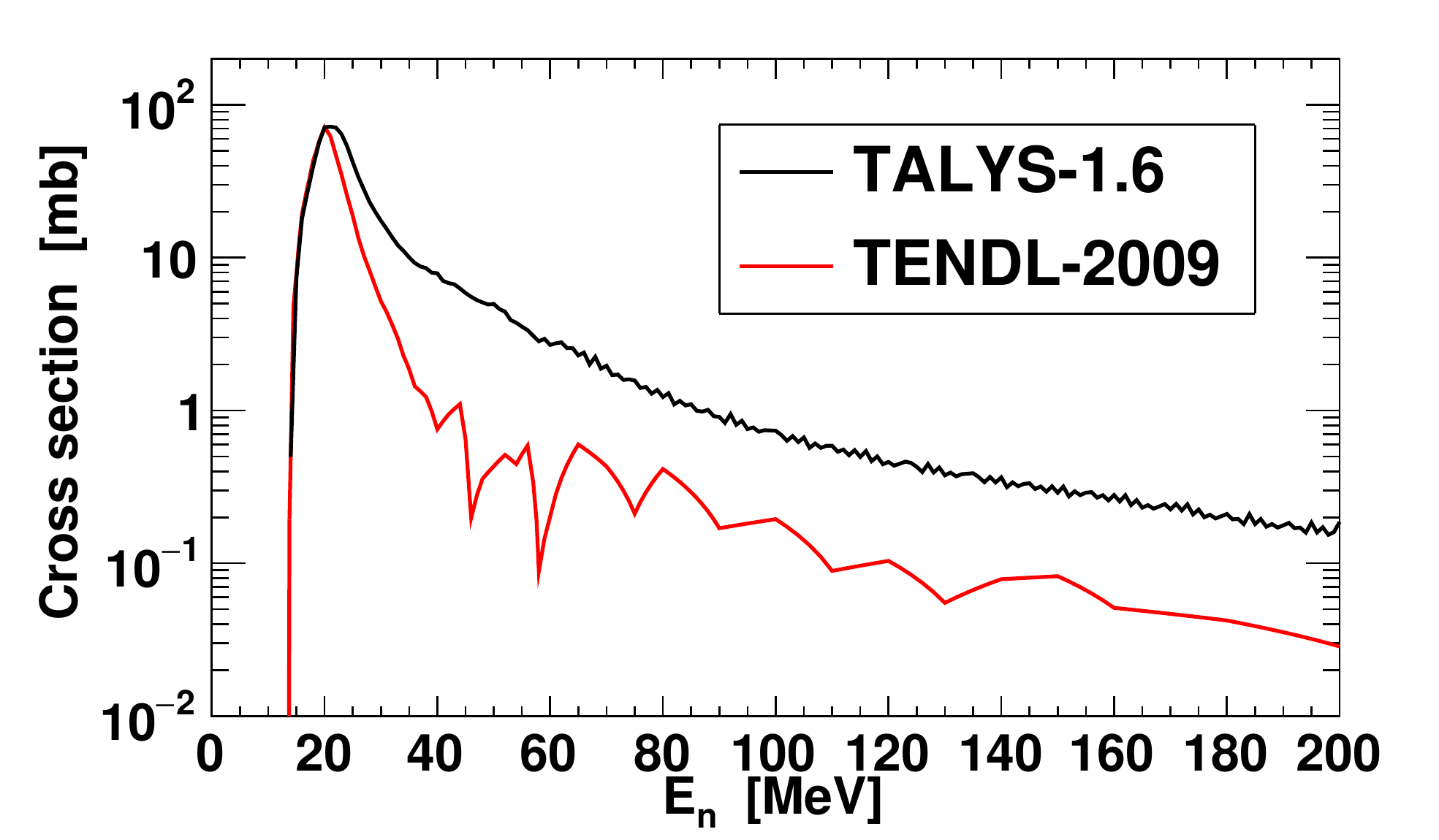}
\caption{$^{12}$C(n,p) cross section computed with TALYS-1.6, in comparison with the data from TENDL-2009 database.}
\label{fig9}
\end{figure}

In the energy range of interest, reactions leading to emission of particles -- such as (n,p), (n,$\alpha$), (n,d) and (n,2n) -- have an important pre-equilibrium component, which can be described either by a semi-classical exciton model, or by a quantum-mechanical multi-step (compound plus direct) model. In the latter option the importance of the multi-step direct (MSD) emission increases quickly with increasing incident energy. The TALYS calculations that closely reproduce the experimentally determined number of produced $^{12}$B nuclei have been obtained by means of a MSD model of Fesbach-Kerman-Koonin type, described in detail in ref.~\cite{Bo94}. The energy dependent cross section from these calculations is shown in fig.~\ref{fig9}. Only one parameter, called \emph{M2constant} in the TALYS user manual \cite{Ko13} and used in the normalization of the distorted-wave Born approximation (DWBA) matrix elements of the model, has been adjusted so as to reproduce the result of the integral measurement at n\_TOF, as will soon be explained. The curve from fig.~\ref{fig9} is obtained with \emph{M2constant}~=~0.45. For comparison, fig.~\ref{fig9} shows also the evaluated cross sections in TENDL-2009, which was based on TALYS calculations with unoptimized parameter. The main difference between the two calculations is above 20 MeV. Below 20 MeV the old and the new TALYS calculations agree between each other and -- as shown in fig.~\ref{fig0} -- with the experimental data near the threshold energy from Kreger and Kern \cite{kreger}, Ablesimov \emph{at al.} \cite{ablesimov} and Bobyr \emph{et al.} \cite{bobyr}. This also goes in the direction of confirming the conclusion by Pillon \emph{et al.} \cite{pillon}, that their cross section may be underestimated due to the difficulties in identifying all possible excited states of the residual nucleus. On the contrary, the peak predicted by the calculations is much higher than the data from Rimmer and Fisher \cite{rimmer}. The old datum at 90~MeV from Kellogg \cite{kellog}, affected by a poor energy resolution, is somewhat underestimated by the computed values. A predicted high-energy tail lower than the few available data appears also in competing reactions -- particularly in the (n,2n) cross section -- and might be due to an oversimplification of the MSD model contained in TALYS-1.6, namely the use of macroscopic form factors in DWBA calculations. Resorting to microscopic form factors based on the shell model would go in the right direction and would improve the high energy behavior. As an alternative, one could try to adjust in an \emph{ad hoc} manner some important parameter, such as the pairing energy. This has not been done in the present work, aiming at reproducing the result of the integral measurement at n\_TOF, rather than the high energy tail of the differential cross section, which can only play a very modest role in this respect.

\begin{table}[t!]
\caption{List of calculated contributions to the number of produced $^{12}$B nuclei and to the quantity analogous to the resonance integral, from different energy regions and from the separate stable carbon isotopes. See the main text for the details of the calculations.}
\label{tab3}
\centering
\begin{tabular}{cc|ccc}
\hline\hline
Quantity&Range&$^{12}$C&$^{13}$C&$^\mathrm{nat}$C\\
\hline
\multirow{3}{*}{$N_{^{12}\mathrm{B}}$}&13.6 MeV -- 200 MeV& 67.50 & 97.09 & 67.83 \\
&200 MeV -- 10 GeV& 0.60 & 65.05 & 1.31 \\
&full range& 68.10 &162.14&\textbf{69.14}\\
\hline
\multirow{3}{*}{\minitab[c]{$I_{^{12}\mathrm{B}}$\\$[$mb$]$}}&13.6 MeV -- 200 MeV&35.68&43.93&35.77\\
&200 MeV -- 10 GeV&0.35&44.51&0.84\\
&full range&36.03&88.44&\textbf{36.61}\\
\hline\hline
\end{tabular}
\end{table}

For the comparison with the n\_TOF integral data, we have computed the contribution $N_{{^{12}\mathrm{B}}}^{(200\:\mathrm{MeV})}$ to the production of $^{12}$B nuclei due to neutrons in the energy range from 13.6~MeV to 200~MeV impinging on $^{12}$C and $^{13}$C, separately, by means of the eq.~(\ref{eq11}), with the weighting function $w(E_\mathrm{n})$ given by the analytical approximation from eq.~(\ref{eq6}). We have also calculated the corresponding contribution \mbox{$\int_{13.6\:\mathrm{MeV}}^{{200\:\mathrm{MeV}}}[\sigma(E_\mathrm{n})/E_\mathrm{n}]\mathrm{d}E_\mathrm{n}$} to the integral analogous to the resonance integral, as defined by eq.~(\ref{eq10}). In case of $^{12}$C as a target, the contribution comes from the (n,p) reaction, while in case of $^{13}$C from (n,np) and (n,d) reactions. The results we have obtained are listed in table~\ref{tab3}. The values for a $^\mathrm{nat}$C target are obtained by a weighted average based on the natural abundances of the two isotopes -- 98.9\% for $^{12}$C and 1.1\% for $^{13}$C.

A final consideration regards the contribution to the two integrals from neutrons in the energy range from 200~MeV to 10~GeV. This has been calculated with version 5.2 of the INCL++ code \cite{Ma14}, where the inclusion of multiple pion production as described in ref.~\cite{Pe11} makes it possible to extend the calculations up to incident nucleon energies of the order of 12 GeV. By default, light nuclei like carbon isotopes undergo the decay known as Fermi break-up. In this high energy range the evaluation of all the cross sections of interest is expected to be quite reliable. The corrections obtained for the $^\mathrm{nat}$C target are also listed in table~\ref{tab3}. The final figures are $N_{^{12}\mathrm{B}}=69.14$, compared to the experimental value $68.5\pm0.4_\mathrm{stat}\pm4.8_\mathrm{syst}$, and $I_{^{12}\mathrm{B}}=36.61$~mb, in comparison with the value of $37\pm3$~mb, as derived from the experiment.

\section{Conclusions}
\label{conclusion}

The integral measurement of the $^{12}$C(n,p)$^{12}$B reaction was performed at the n\_TOF facility at CERN, from the reaction threshold at 13.6~MeV, up to 10~GeV. The measurement was performed using two C$_6$D$_6$ detectors, commonly used at n\_TOF for neutron capture measurements. The high energy of $\beta$-rays coming from the decay of $^{12}$B nuclei produced by neutron activation of $^{12}$C, allows them to reach the scintillator and deposit inside it a large amount of energy. The high instantaneous neutron flux and the low repetition rate of the n\_TOF beam make the n\_TOF facility well suited for such activation measurements, especially when the decaying nuclides are characterized by a half-life below $\sim$100~ms. The measurements at n\_TOF are affected by several sources of background. Most of them -- the background of scattered in-beam $\gamma$-rays, the one related to the neutron beam crossing the experimental area and the ambient background -- have been measured and subtracted. The remaining background component -- caused by the neutrons scattering off the sample itself -- has been studied by means of recently developed dedicated simulations.

After subtracting all background components from the $^\mathrm{nat}$C measurements, the exponential spectra perfectly reproduce the expected lifetime of $^{12}$B. After correction for the efficiency, a fit of the time spectra to the exponential form yields the number $N_{^{12}\mathrm{B}}$ of $^{12}$B nuclei produced per neutron bunch. The results from the two C$_6$D$_6$ detectors were found to be highly consistent, yielding the final result of $N_{^{12}\mathrm{B}}=68.5\pm0.4_\mathrm{stat}\pm4.8_\mathrm{syst}$. The n\_TOF result has been compared with evaluated cross sections and model calculations, in particular those used in GEANT4. In all cases cross sections were folded with the n\_TOF neutron flux and corrected for self-shielding and multiple scattering effects. The neutron flux up to 1~GeV was measured in the past by  the Parallel Plate Avalanche Counters (PPAC), relying on the $^{235}$U(n,f) reaction, whose cross section is considered a standard up to 200~MeV. Above 200~MeV the neutron flux has been reevaluated due to the recent extension of the cross section reference for this reaction. The reevaluation of the experimental data has also affected the selection of the simulated results used to extend the evaluation of the neutron flux up to 10~GeV.

For most evaluations the cross section of the $^{12}$C(n,p)$^{12}$B reaction is underestimated, as they are based on the data of Rimmer \emph{et al.} Among models in GEANT4, good agreement is observed only with a combined Bertini-Binary cascade model. A comparison was also performed with theoretical cross sections obtained by the TALYS-1.6 code, whose parameters were optimized to reproduce the integral n\_TOF value. Since the theoretical model predicts the integral value of the cross section, as well as its full energy dependence, the comparison with the n\_TOF result can be used to extract the energy dependent cross section up to 10 GeV.

The models and evaluations that yield the integral value closest to the one from n\_TOF indicate that the cross section for the $^{12}$C(n,p)$^{12}$B reaction reaches its maximum around 20 MeV. The few experimental data that cover this energy range also seem to confirm this observation. However, many of the evaluations based on these data underestimate the integral value of the cross section, relative to the one from n\_TOF. Therefore, a renewed measurement of the energy dependence of this cross section is strongly encouraged by the latest n\_TOF result, in particular within the energy range from the  reaction threshold up to several tens of MeV.\\

This research was funded by the European Atomic Energy Community’s (Euratom) Seventh Framework Programme FP7/2007-2011 under the Project CHANDA (Grant No. 605203); by the Narodowe Centrum Nauki (NCN) grant UMO-2012/04/M/ST2/00700; and by the Croatian Science Foundation under Project No. 1680. We are grateful to Drs. Arjan J. Koning and Davide Mancusi for valuable comments on the use of TALYS-1.6 and INCL++v5.2, respectively. GEANT4 simulations have been run at the Laboratory for Advanced Computing, Faculty of Science, University of Zagreb.


\end{document}